\documentclass[unsortedaddress,nopreprint,twocolumn,numbers]{jasatex}

 \usepackage{amssymb, latexsym}
 \usepackage{amsmath} 
 \usepackage{amsthm}
 \usepackage{bm}
 \usepackage{graphicx}
 \usepackage{epstopdf}
 \usepackage[mathscr]{eucal}
 \usepackage{enumerate}
 \usepackage{listings}

\begin{document}

\title[Grid-free compressive beamforming]{Grid-free compressive beamforming}

\author{Angeliki Xenaki}
\email[Author to whom correspondence should be addressed. Electronic mail: ]{anxe@dtu.dk}
\affiliation{Department of Applied Mathematics and Computer Science, Technical University of Denmark, Kgs.Lyngby, 2800 Denmark}
\author{Peter Gerstoft}
\affiliation{Scripps Institution of Oceanography, University of California San Diego, La Jolla, California 92093--0238}

\date{\today}

\begin{abstract}
The direction-of-arrival (DOA) estimation problem involves the localization of a few sources from a limited number of observations on an array of sensors, thus it can be formulated as a sparse signal reconstruction problem and solved efficiently with compressive sensing (CS) to achieve high-resolution imaging. On a discrete angular grid, the CS reconstruction degrades due to basis mismatch when the DOAs do not coincide with the angular directions on the grid. To overcome this limitation, a continuous formulation of the DOA problem is employed and an optimization procedure is introduced, which promotes sparsity on a continuous optimization variable. The DOA estimation problem with infinitely many unknowns, i.e., source locations and amplitudes, is solved over a few optimization variables with semidefinite programming. The grid-free CS reconstruction provides high-resolution imaging even with non-uniform arrays, single-snapshot data and under noisy conditions as demonstrated on experimental towed array data. 
\end{abstract}

\pacs{
43.60.Pt, 43.60.Jn, 43.60.Fg
}

\maketitle

\section{\label{sec:Intro}Introduction} 
 
Sound source localization with sensor arrays involves the estimation of the direction-of-arrival (DOA) of (usually a few) sources from a limited number of observations. Compressive sensing\cite{EladBook:2010, FoucartBook:2013} (CS) is a method for solving such underdetermined problems with a convex optimization procedure which promotes sparse solutions.

Solving the DOA estimation as a sparse signal reconstruction problem with CS, results in robust, high-resolution acoustic imaging\cite{MalioutovDOA:2005,EdelmannCSDOA:2011,XenakiCS:2014,Mecklenbrauker:2013}, outperforming traditional methods\cite{KrimDOA:1996} for DOA estimation. Furthermore, in ocean acoustics, CS is shown to improve the performance of matched field processing\cite{MantzelCMFP:2012,Forero:2014}, which is a generalized beamforming method for localizing sources in complex environments (e.g., shallow water), and of 
coherent passive fathometry in inferring the number and depth of sediment layer interfaces\cite{YardimCSfathometer:2014}.

One of the limitations of CS in DOA estimation is basis mismatch\cite{ChiBasisMismatch:2011} which occurs when the sources do not coincide with the look directions due to inadequate discretization of the angular spectrum. Under basis mismatch, spectral leakage leads to inaccurate reconstruction, i.e., estimated DOAs deviating from the actual ones. Employing finer grids\cite{MalioutovDOA:2005, DuarteBasisMismatch:2013} alleviates basis mismatch at the expense of increased computational complexity, especially in large two-dimensional or three-dimensional problems as encountered in seismic imaging for example\cite{YaoCSseismic:2011,YaoCSseismic:2013, Fan:2014}.

To overcome basis mismatch, we formulate the DOA estimation problem in a continuous angular spectrum and introduce a sparsity promoting measure for general signals, the atomic norm\cite{Chandrasekaran2012}. The atomic norm minimization problem, which has infinitely many unknows, is solved efficiently over few optimization variables in the dual domain with semidefinite programming\cite{GrandaSuperRes2014}. Utilizing the dual optimal variables, we show that the DOAs are accurately reconstructed through polynomial rooting. It is demonstrated that grid-free CS gives robust, high-resolution reconstruction also with non-uniform arrays and noisy measurements, exhibiting great flexibility in practical applications.

Polynomial rooting is employed in several DOA estimation methods to improve the resolution. However, these methods involve the estimation of the cross-spectral matrix hence they require many snapshots and stationary incoherent sources and are suitable only for uniform linear arrays (ULA)\cite{VanTreesBook}. Grid-free CS is demonstrated not to have these limitations.

Finally, we process acoustic data\cite{FORAdata:2011} from measurements in the North-East (NE) Pacific with grid-free CS and demonstrate that the method provides high-resolution acoustic imaging even with single-snapshot data.

In this paper, vectors are represented by bold lowercase letters and matrices by bold uppercase letters. The symbols $^{T}$, $^{H}$ denote the transpose and the Hermitian (i.e., conjugate transpose) operator respectively on vectors and matrices. The symbol $^{*}$ denotes simple conjugation. The generalized inequality $\mathbf{X}\succeq 0$ denotes that the matrix $\mathbf{X}$ is positive semidefinite. The $\ell_{p}$-norm of a vector $\mathbf{x} \in\mathbb{C}^{n}$ is defined as $\lVert\textbf{x}\rVert_{p} = \left(\sum_{i=1}^{n}\lvert x_{i}\rvert^{p}\right)^{1/p}$. By extension, the $\ell_{0}$-norm is defined as $\lVert\textbf{x}\rVert_{0}=\sum_{i=1}^{n} 1_{x_{i}\neq0}$. The paper makes heavy use of convex optimization theory; for a summary see App.~\ref{sec:AppOptimizationProblems}.

\section{\label{sec:DiscreteDOA}Discrete DOA estimation}

The DOA estimation problem involves the localization of usually a few sources from measurements on an array of sensors. For simplicity, we assume that the sources are in the far-field of the array, such that the wavefield impinging on the array consists of a superposition of plane waves, that the processing is narrowband and the sound speed is known. Moreover, we consider the one-dimensional problem with a uniform linear array of sensors and the sources residing in the plane of the array.

The location of a source is characterized by the direction of arrival of the associated plane wave, $\theta \in [-90^{\circ}, 90^{\circ}]$, with respect to the array axis. The propagation delay from the $i$th potential source to each of the $M$ array sensors is described by the steering (or replica) vector,
\begin{equation}
\mathbf{a}(\theta_{i})=e^{j 2\pi\frac{d}{\lambda}\left[ 0,\cdots, M-1\right]^{T}\sin\theta_{i}},
\label{eq:SteeringVector} 
\end{equation}
\noindent where $\lambda$ is the wavelength and $d$ is the intersensor spacing.

Discretizing the half-space of interest, $\theta \in [-90^{\circ}, 90^{\circ}]$, into $N$ angular directions the DOA estimation problem is expressed in a matrix-vector formulation,
\begin{equation}
\mathbf{y} = \mathbf{A}\mathbf{x},
\label{eq:DiscreteDOAsensor}
\end{equation}
\noindent where $\mathbf{y}\in \mathbb{C}^{M}$ is the vector of the wavefield measurements at the $M$ sensors, $\mathbf{x}\in \mathbb{C}^{N}$ is the unknown vector of the complex source amplitudes at all $N$ directions on the angular grid of interest and $\mathbf{A}$ is the sensing matrix which maps the signal to the observations,
\begin{equation}
\mathbf{A}_{M\times N}=[\mathbf{a}(\theta_{1}), \cdots, \mathbf{a}(\theta_{N})].
\label{eq:SensingMatrix}
\end{equation}
In the presence of additive noise $\mathbf{n}\in\mathbb{C}^{M}$, the measurement vector is described by,
\begin{equation}
\mathbf{y} = \mathbf{A}\mathbf{x} + \mathbf{n}.
\label{eq:NoisyData}
\end{equation}
The noise is generated as independent and identically distributed (iid) complex Gaussian. The array signal-to-noise ratio (SNR) for a single-snapshot is used in the simulations, defined as SNR=$20\log_{10}\left(\lVert\mathbf{A}\mathbf{x}\rVert_{2}/\lVert\mathbf{n}\rVert_{2}\right)$, which determines the noise $\ell_{2}$-norm, $\lVert\mathbf{n}\rVert_{2} = \lVert\mathbf{A}\mathbf{x}\rVert_{2} 10^{-\mathrm{SNR}/20}$.

\subsection{\label{sec:BeamformingCS}Sparse signal reconstruction}

Practically, we are interested in a fine resolution on the angular grid such that $M<N$ and the problem~\eqref{eq:DiscreteDOAsensor} is underdetermined. A way to solve this ill-posed problem is to constrain the possible solutions with prior information.

Traditional methods solve the underdetermined problem~\eqref{eq:DiscreteDOAsensor} by seeking the solution with the minimum $\ell_{2}$-norm which fits the data as described by the minimization problem,
\begin{equation}
\underset{\mathbf{x}\in\mathbb{C}^{N}}{\text{min}}\lVert\mathbf{x}\rVert_{2} \; \text{subject to} \;  \mathbf{y}=\mathbf{A}\mathbf{x}.
\label{eq:CS_l2}
\end{equation}
\noindent The minimization problem~\eqref{eq:CS_l2} is convex with analytic solution, $\mathbf{\hat{x}} = \mathbf{A}^{H}\left(\mathbf{A}\mathbf{A}^{H}\right)^{-1}\mathbf{y}$. However, it aims to minimize the energy of the signal rather than its sparsity, hence the resulting solution is non-sparse. 

Conventional beamforming\cite{JohnsonBook1993} (CBF) is the simplest source localization method and it is based on the $\ell_{2}$-norm method with the simplifying condition $\mathbf{A}\mathbf{A}^{H} = \mathbf{I}_{M}$. CBF combines the sensor outputs coherently to enhance the signal at a specific look direction from the ubiquitous noise yielding the solution,
\begin{equation}
\mathbf{\hat{x}}_{\text{CBF}} = \mathbf{A}^{H}\mathbf{y}.
\label{eq:CBF}
\end{equation}
\noindent CBF is robust to noise but suffers from low resolution and the presence of sidelobes.

A sparse solution $\mathbf{x}$ is preferred by minimizing the $\ell_{0}$-norm leading to the minimization problem,
\begin{equation}
\underset{\mathbf{x}\in\mathbb{C}^{N}}{\text{min}}\lVert\mathbf{x}\rVert_{0} \; \text{subject to} \;  \mathbf{y}=\mathbf{A}\mathbf{x}.
\label{eq:CS_l0}
\end{equation}
However, the minimization problem~\eqref{eq:CS_l0} is a non-convex combinatorial problem which becomes computationally intractable even for moderate dimensions. The breakthrough of compressive sensing\cite{EladBook:2010, FoucartBook:2013} (CS) came with the proof that for sufficiently sparse signals, $K<<N$, $K<M$, and sensing matrices with sufficiently incoherent columns the minimization problem~\eqref{eq:CS_l0} is equivalent to the minimization problem,
\begin{equation}
\underset{\mathbf{x}\in\mathbb{C}^{N}}{\text{min}}\lVert\mathbf{x}\rVert_{1} \; \text{subject to} \;  \mathbf{y}=\mathbf{A}\mathbf{x},
\label{eq:CS_l1}
\end{equation}
\noindent where the $\ell_{0}$-norm is replaced with the $\ell_{1}$-norm. The problem~\eqref{eq:CS_l1} is the closest convex optimization problem to the problem~\eqref{eq:CS_l0} and can be solved efficiently by convex optimization even for large dimensions\cite{BaraniukCSNotes:2007}.

For noisy measurements~\eqref{eq:NoisyData}, the constraint in~\eqref{eq:CS_l1} becomes $\lVert \mathbf{y} - \mathbf{A}\mathbf{x} \rVert_{2}\leq\epsilon$, where $\epsilon$ is the noise floor, i.e., $\lVert\mathbf{n}\rVert_{2}\leq \epsilon$. Then, the solution is\cite{TroppCSNoise:2006}, 
\begin{equation}
\hat{\mathbf{x}}_{\text{CS}}=\underset{\mathbf{x}\in\mathbb{C}^{N}}{\text{argmin}}\lVert\mathbf{x}\rVert_{1} \; \text{subject to} \;\lVert \mathbf{y} - \mathbf{A}\mathbf{x} \rVert_{2}\leq\epsilon,
\label{eq:CS_solution_noisy}
\end{equation}
\noindent which has the minimum $\ell_{1}$-norm while it fits the data up to the noise level.

Herein, we use the cvx toolbox for disciplined convex optimization which is available in the Matlab environment. It uses interior point solvers to obtain the global solution of a well-defined optimization problem\cite{CVX}. Interior point methods solve an optimization problem with linear equality and inequality constraints by transforming it to a sequence of simpler linear equality constrained problems which are solved iteratively with the Newton's method  (iterative gradient descent method) increasing the accuracy of approximation at each step\cite{BoydBook}.

\subsection{\label{sec:BasisMismatch}Basis Mismatch}

CS offers improved resolution due to the sparsity constraint and it can be solved efficiently with convex optimization. However, CS performance in DOA estimation is limited by the coherence of the sensing matrix $\mathbf{A}$ (see Ref. \onlinecite{XenakiCS:2014}), described by the restricted isometry property\cite{CandesRIP:2008}, and by basis mismatch\cite{ChiBasisMismatch:2011,DuarteBasisMismatch:2013} due to inadequate discretization of the angular grid. Herein, we demonstrate a way to overcome the limitation of basis mismatch by solving the $\ell_{1}$-minimization problem on a grid-free, continuous spatial domain.

\begin{figure}[tb]
\centering
\includegraphics[width=8.6cm]{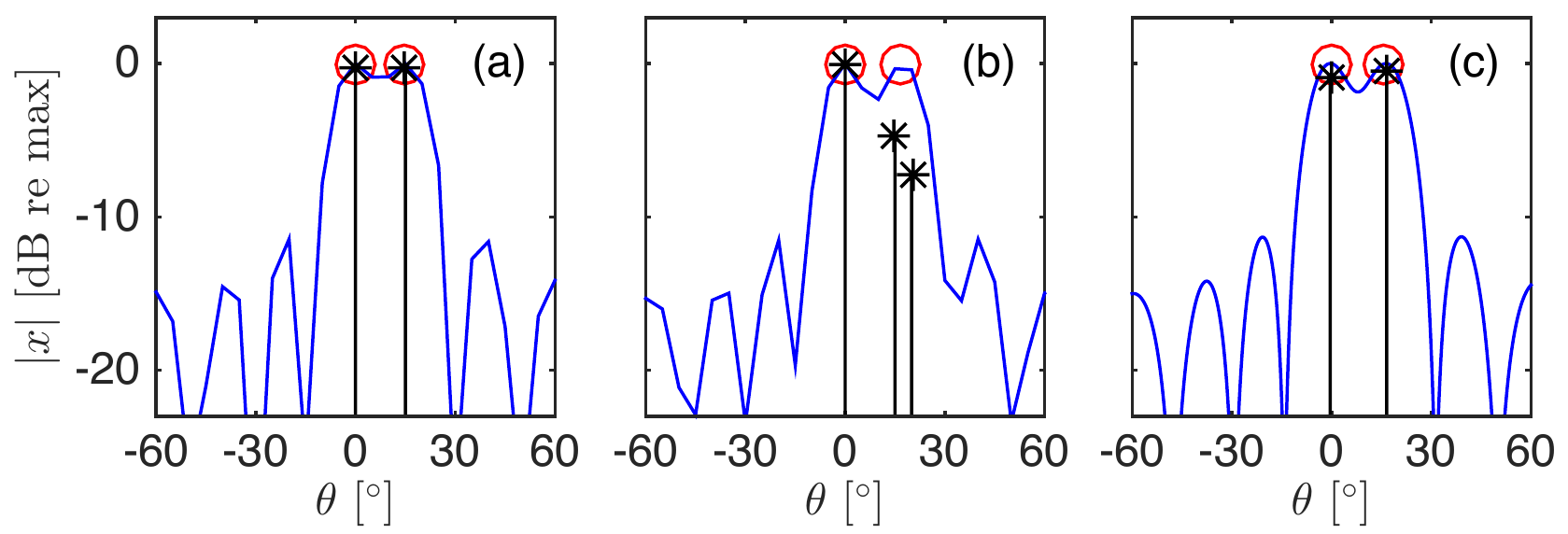}
\caption{(Color online) CS performance in DOA estimation in terms of the discretization of the angular space. A standard ULA is used with $M$=$8$ sensors, $d / \lambda = 1 / 2$ and SNR = $20$~dB. CBF and CS (*) reconstruction of two sources (o) (a) at $0^{\circ}$ and $15^{\circ}$ on a grid [$-90^{\circ}$:$5^{\circ}$:$90^{\circ}$], (b) at $0^{\circ}$ and $17^{\circ}$ on a grid [$-90^{\circ}$:$5^{\circ}$:$90^{\circ}$] and (c) at $0^{\circ}$ and $17^{\circ}$ on a grid [$-90^{\circ}$:$1^{\circ}$:$90^{\circ}$].}
\label{fig:BasisMismatch}
\end{figure}

The fundamental assumption in CS is the sparsity of the underlying signal  in the basis of representation, i.e., the sensing matrix $\mathbf{A}$. However, when the sources do not match with the selected angular grid, the signal might not appear sparse in the selected DFT basis~\cite{ChiBasisMismatch:2011}. Figure~\ref{fig:BasisMismatch} shows the degradation of CS performance under basis mismatch due to inadequate discretization of the DOA domain in FFT beamforming.

To increase the precision of the CS reconstruction, Malioutov \textit{et al.}\cite{MalioutovDOA:2005} and Duarte and Baraniuk\cite{DuarteBasisMismatch:2013} propose an adaptive grid refinement. The adaptive grid refinement aims at improving the resolution of CS reconstruction without significant increase in the computational complexity by first detecting the regions where sources are present on a coarse grid and then refining the grid locally only at these regions. Grid refinement is an intuitive way of circumventing basis mismatch. However, the problem of basis mismatch is avoided only if the problem is solved in a continuous setting, particularly for moving sources.  

\section{\label{sec:ContinuousDOA}Continuous DOA estimation}

In the continuous approach, the $K$-sparse signal, $x$, is expressed as,
\begin{equation}
x(t) = \sum _{i=1}^{K} x_{i} \delta(t-t_{i}),
\label{eq:SuperpositionOfSpikes}
\end{equation}
\noindent where $x_{i}\in \mathbb{C}$ is the complex amplitude of the $i$th source, $t_{i}=\sin\theta_{i}$ is its support, i.e, the corresponding DOA, on the continuous sine spectrum $\mathbb{T}=[-1,1]$ (with $T\subset\mathbb{T}$ the set of the DOAs of all $K$ sources) and $\delta(t)$ is the Dirac delta function.


The sound pressure received at the $m$th sensor is expressed as a superposition of plane waves from all possible directions on the continuous sine spectrum $\mathbb{T}$, 
\begin{equation}
y_{m} = \int\limits_{-1}^{1} \! x(t)e^{j2\pi \frac{d}{\lambda}(m-1)t} \,\mathrm{d}t = \sum\limits_{i=1}^{K} x_{i}e^{j2\pi \frac{d}{\lambda}(m-1)t_{i}},
\label{eq:SensorPressure}
\end{equation}
\noindent and the measurement vector of the sensor array is,
\begin{equation}
\mathbf{y}_{M\times 1} = \mathcal{F}_{M}x,
\label{eq:MeasurementVector}
\end{equation}
\noindent where $\mathcal{F}_{M}$ is a linear operator (inverse Fourier transform) which maps the continuous signal $x$ to the observations $\mathbf{y}\in \mathbb{C}^{M}$.

In the presence of additive noise, $\mathbf{n}\in\mathbb{C}^{M}$, the measurement vector is described by,
\begin{equation}
\mathbf{y} = \mathcal{F}_{M}x + \mathbf{n},
\label{eq:NoisyDataContinuous}
\end{equation}
\noindent similarly to Eq.~\eqref{eq:NoisyData}.

\section{\label{sec:NoiselessReconstruction}Grid-free sparse reconstruction}

To solve the underdetermined problem~\eqref{eq:MeasurementVector} (or equivalently the problem~\eqref{eq:NoisyDataContinuous}) in favor of sparse solutions, we describe an optimization procedure which promotes sparsity on a continuous optimization variable.

\subsection{Atomic norm}

In the discrete formulation~\eqref{eq:NoisyData} of the DOA estimation problem, the prior information about the sparse distribution of sources is imposed through the $\ell_{1}$-norm of the vector $\mathbf{x}$ to obtain sparse estimates~\eqref{eq:CS_solution_noisy}. By extension, in the continuous formulation~\eqref{eq:NoisyDataContinuous}, we introduce the atomic norm\cite{Chandrasekaran2012}, $\lVert\cdot\rVert_{\mathcal{A}}$, as a sparsity promoting measure for the continuous signal $x(t)$ in Eq.~\eqref{eq:SuperpositionOfSpikes} defined as,
\begin{equation}
\lVert x\rVert_{\mathcal{A}} = \sum\limits_{i=1}^{K} \lvert x_{i}\rvert.
\label{eq:AtomicNorm}
\end{equation}
\noindent In other words, the atomic norm is a measure for continuous signals equivalent to the $\ell_{1}$-norm (which is defined only on vector spaces). Hence, the atomic norm is a convex function which promotes sparsity in a general framework. For a discrete grid the atomic norm corresponds to the $\ell_{1}$-norm.

To clarify the analogy between the $\ell_{1}$-norm and the atomic norm and justify the term \textit{atomic}, consider that the vector $\mathbf{x}\in\mathbb{C}^{N}$ can be interpreted as a linear combination of $N$ unit vectors. The unit vectors, in this case, are the smallest units, or \textit{atoms}, in which the vector $\mathbf{x}$ can be decomposed into. The $\ell_{1}$-norm is the sum of the absolute values of the weights of this linear combination of atoms\cite{BoydBook}.

Analogously, the continuous signal~\eqref{eq:SuperpositionOfSpikes} can be interpreted as a linear combination of $K$ delta functions $\delta(t-t_{i})$, serving as atoms for the continuous signal $x(t)$ and the atomic norm is the sum of the absolute values of the weights of the linear combination of these atoms\cite{Chandrasekaran2012}. Even though there are infinitely many atoms in the continuous case, only few of those, $K<M$,  constitute the signal and the sum in~\eqref{eq:AtomicNorm} is finite.

\subsection{Primal problem}

Utilizing the convex measure of the atomic norm, the DOA estimation in the continuous angular space is solved with the sparsity promoting minimization problem,
\begin{equation}
\underset{x}{\min}\lVert x\rVert_{\mathcal{A}} \; \text{subject to} \; \mathbf{y}=\mathcal{F}_{M}x.
\label{eq:PrimalProblem}
\end{equation}

Since the optimization variable $x$ is a continuous parameter, the primal problem~\eqref{eq:PrimalProblem} is infinite dimensional and cannot be solved as such. It is possible to approximate the continuous variable $x$ on a discrete grid and solve the $\ell_{1}$-norm optimization problem~\eqref{eq:CS_l1}. This would increase the computational complexity significantly when the discretization step is reduced to improve precision. An alternative to this, is to gradually refine the discretization step\cite{MalioutovDOA:2005}. However, we show that by solving the dual problem instead, there is no need to employ a discrete approximation of the continuous variable, $x$.

\subsection{\label{sec:DualProblem}Dual problem}

To formulate the dual problem to the problem~\eqref{eq:PrimalProblem} (see Appendix~\ref{sec:AppOptimizationProblems} for details), we construct the Lagrangian by making the explicit equality constraints, $\mathbf{y} =\mathcal{F}_{M}x$, implicit in the objective function,
\begin{equation}
L(x,\mathbf{c}) = \lVert x\rVert_{\mathcal{A}} + \mathrm{Re}\left[ \mathbf{c}^{H}\left( \mathbf{y}-\mathcal{F}_{M}x\right)\right],
\label{eq:AtomicLagrangian}
\end{equation}
\noindent where $\mathbf{c}\in\mathbb{C}^{M}$ is the vector of dual variables.

The dual function $g(\mathbf{c})$ is the infimum, i.e., the greatest lower bound, of the Lagrangian, $L(x,\mathbf{c})$, over the primal optimization variable $x$,
\begin{equation}
\begin{aligned}
g(\mathbf{c}) &= \underset{x}{\inf} \; L(x,\mathbf{c})\\ 
&= \mathrm{Re}\left[ \mathbf{c}^{H}\mathbf{y}\right] + \underset{x}{\inf} \; \left( \lVert x\rVert_{\mathcal{A}} - \mathrm{Re}\left[\mathbf{c}^{H}\mathcal{F}_{M}x\right] \right).
\end{aligned}
\label{eq:AtomicDualFunction}
\end{equation}

To evaluate the second term in~\eqref{eq:AtomicDualFunction} we note that for every $x_{i}$, $\mathrm{Re}\left[\left(\mathbf{c}^{H}\mathcal{F}_{M}\right)_{i} x_{i}\right] = \mathrm{Re}\left[\left(\mathcal{F}_{M}^{H}\mathbf{c}\right)_{i}^{H} x_{i}\right] = \lvert\left(\mathcal{F}_{M}^{H}\mathbf{c}\right)_{i}\rvert\lvert x_{i}\rvert\cos\phi_{i}$, where $\phi_{i}$ is the angle between $x_{i}$ and $\left(\mathcal{F}_{M}^{H}\mathbf{c}\right)_{i}$. Then,
\begin{equation}
\begin{aligned}
\lvert x_{i} \rvert \! - \! \mathrm{Re}\left[\left(\mathcal{F}_{M}^{H}\mathbf{c}\right)_{i}^{H}\!x_{i}\right] \! &= \! \lvert x_{i}\rvert \!\left[ 1 \! - \! \lvert\left(\mathcal{F}_{M}^{H}\mathbf{c}\right)_{i}\rvert \cos\phi_{i}\right]\\
& \geq \lvert x_{i}\rvert \!\left[ 1 \! - \! \lvert\left(\mathcal{F}_{M}^{H}\mathbf{c}\right)_{i}\rvert\right].
\end{aligned}
\label{eq:InfimumDualFunction}
\end{equation}
\noindent The lower bound in~\eqref{eq:InfimumDualFunction} is nonnegative if $\lvert\mathcal{F}_{M}^{H}\mathbf{c}\rvert$ is less than one, $\max_{i}\lvert\left(\mathcal{F}_{M}^{H}\mathbf{c}\right)_{i}\rvert\leq 1$, and the infimum is zero. Otherwise, $\lvert x_{i}\rvert \!\left[ 1 \! - \! \lvert\left(\mathcal{F}_{M}^{H}\mathbf{c}\right)_{i}\rvert\right] < 0$ and the infimum is attained at $-\infty$. Hence, the dual function is,
\begin{equation}
\begin{aligned}
g(\mathbf{c}) = \left\{ \begin{array}{rl}
& \mathrm{Re}\left[ \mathbf{c}^{H}\mathbf{y}\right], \quad \lVert\mathcal{F}_{M}^{H}\mathbf{c}\rVert_{\infty}\leq 1 \\
& -\infty, \quad\quad\quad\quad \text{otherwise}.
\end{array} \right.
\end{aligned}
\label{eq:AtomicDualFunction2}
\end{equation}

From~\eqref{eq:InfimumDualFunction}, $\lvert x_{i}\rvert \left[ 1 - \lvert\left(\mathcal{F}_{M}^{H}\mathbf{c}\right)_{i}\rvert \cos\phi_{i}\right]=0$ at the infimum, which for every $x_{i}\neq 0$ yields $\lvert\left(\mathcal{F}_{M}^{H}\mathbf{c}\right)_{i}\rvert\cos\phi_{i}=1$, i.e., $\lvert\left(\mathbf{c}^{H}\mathcal{F}_{M}\right)_{i}\rvert=1$  and $\phi_{i} = 0$, as both $\lvert\left(\mathbf{c}^{H}\mathcal{F}_{M}\right)_{i}\rvert\leq 1$ and $\cos\phi_{i}\leq 1$. Thus, for $x_{i}\neq 0$, $\left( \mathcal{F}_{M}^{H}\hat{\mathbf{c}}\right)_{i}$ is a unit vector in the direction of $x_{i}$,
\begin{equation}
\begin{aligned}
&\left( \mathcal{F}_{M}^{H}\hat{\mathbf{c}}\right)_{i} = x_{i}/\lvert x_{i}\rvert, \;   x_{i}\neq 0 \\
&\lvert \mathcal{F}_{M}^{H}\hat{\mathbf{c}}_{i} \rvert <1, \;\quad\quad x_{i}= 0.
\end{aligned}
\label{eq:DualConstraint}
\end{equation}

Maximizing the dual function~\eqref{eq:AtomicDualFunction2} constitutes the dual problem,
\begin{equation}
\underset{\mathbf{c}\in\mathbb{C}^{M}}{\max}\;\mathrm{Re}\left[\mathbf{c}^{H}\mathbf{y}\right] \; \text{subject to} \;  \lVert\mathcal{F}^{H}_{M}\mathbf{c}\rVert_{\infty}\leq 1.
\label{eq:DualProblem}
\end{equation}
Since the primal problem~\eqref{eq:PrimalProblem} is convex with linear equality constraints~\eqref{eq:SlatersCondition}, strong duality holds assuring that the maximum of the dual problem~\eqref{eq:DualProblem} is equal to the minimum of the primal problem.

The dual problem~\eqref{eq:DualProblem} selects a vector $\mathbf{c}\in\mathbb{C}^{M}$ which is maximally aligned with the measurement vector $\mathbf{y}\in\mathbb{C}^{M}$ while its beamformed amplitude $\lvert\mathcal{F}^{H}_{M}\mathbf{c}\rvert$ is bounded by unity across the whole angular spectrum. At the angular direction corresponding to the DOA of an existing source, the beamformed dual vector~\eqref{eq:DualConstraint} is equal to the normalized source amplitude.

\subsection{Dual problem using semidefinite programming}

The dual problem \eqref{eq:DualProblem} is a semi-infinite programming problem with a finite number of optimization variables, $\mathbf{c}\in \mathbb{C}^{M}$, and infinitely many inequality constraints, which is still intractable.

Define the dual polynomial,
\begin{equation}
H(z) = \mathcal{F}^{H}_{M}\mathbf{c} = \sum\limits_{m=0}^{M-1} c_{m} z^{m}=\sum\limits_{m=0}^{M-1} c_{m} e^{-j\left(2\pi\frac{d}{\lambda}t\right) m}.
\label{eq:DualPolynomialt}
\end{equation}
\noindent Note that $\mathcal{F}_{M}^{H}\mathbf{c}$ is a trigonometric polynomial~\eqref{eq:TrigonometricPolynomial}, of the variable $z(t) = e^{-j2\pi\frac{d}{\lambda}t}$, $t\in\mathbb{T}$, with the dual variables $\mathbf{c}=[c_{0},\cdots,c_{M-1}]^{T}$ as coefficients and degree $M-1$.

The inequality constraint in Eq.~\eqref{eq:DualProblem} implies that the dual polynomial has amplitude uniformly bounded for all $t\in\mathbb{T}$; see Eq.~\eqref{eq:BoundedLinfty}. Making use of the approximation in Eq.~\eqref{eq:BRL} for bounded trigonometric polynomials, the constraint in Eq.~\eqref{eq:DualProblem} can be replaced with finite dimensional linear matrix inequalities. Thus, the dual problem is solved with semidefinite programming\cite{BoydBook, CVX}, i.e., a convex optimization problem where the inequality constraints are linear matrix inequalities with semidefinite matrices,
\begin{equation}
\begin{aligned}
\underset{\mathbf{c},\mathbf{Q}}{\max}\;\mathrm{Re}\left(\mathbf{c}^{H}\mathbf{y}\right) \; \text{subject to} \begin{bmatrix} \mathbf{Q}_{M\times M} & \mathbf{c}_{M\times 1}\\
\mathbf{c}^{H}_{1\times M} & 1\end{bmatrix} \succeq 0,&\\
\sum\limits_{i=1}^{M-j} \mathbf{Q}_{i,i+j} = \left\{ \begin{array}{rl}
 1, &j=0 \\
 0, &j=1,\cdots,M-1.
       \end{array} \right.&
\end{aligned}
\label{eq:DualProblemSP}
\end{equation}

The number of optimization variables of the dual problem~\eqref{eq:DualProblemSP} is $(M+1)^2/2$ equal to half the number of elements of the Hermitian matrix in the inequality constraint. Thus, a problem with infinitely many unknown parameters~\eqref{eq:PrimalProblem} is solved over a few optimization variables.

\subsection{\label{sec:SupportDetection}Support detection through the dual polynomial}

Strong duality assures that by solving the dual problem~\eqref{eq:DualProblem}, or equivalently Eq.~\eqref{eq:DualProblemSP}, we obtain the minimum of the primal problem~\eqref{eq:PrimalProblem}. However, the dual problem provides an optimal dual vector, $\hat{\mathbf{c}}$, but not the primal solution, $\hat{x}$. Since the corresponding dual polynomial, $H(z) = \mathcal{F}^{H}_{M}\hat{\mathbf{c}}$, has the properties in Eq.~\eqref{eq:DualConstraint}, the support $\hat{T}$ of the primal solution $\hat{x}$ can be estimated by locating the angular directions $t_{i}$ where the amplitude of the dual polynomial is one (i.e., the angular directions at the maxima of the beamformed dual vector),
\begin{equation}
\lvert H(z) \rvert \leq 1, \; \forall t\in \mathbb{T} \rightarrow
\left\{ \begin{array}{rl}
&\!\!\!\! \lvert H\left(z(t_{i})\right)\lvert = 1, \; t_{i}\in \hat{T} \\
&\!\!\!\! \lvert H\left(z(t)\right)\rvert< 1, \; t\in\mathbb{T}\backslash \hat{T}.
       \end{array} \right.
\label{eq:DualPolyAmplitude}
\end{equation}

Following Sec.~\ref{sec:Roots}, this is done by locating the roots of the nonnegative polynomial which lie on the unit circle $\lvert z\rvert=1 $ (see also Sec.~\ref{sec:RootDOA}),
\begin{equation}
P(z) = 1-R(z) = 1 - \sum\limits_{m=-(M-1)}^{M-1} r_{m} z^{m},
\label{eq:RootPoly}
\end{equation}
\noindent where $R(z) = H(z)H(z)^{H}= \lvert H(z) \rvert^{2}$ with coefficients $r_{m} = \sum_{l=0}^{M-1-m} \hat{c}_{l}\hat{c}_{l+m}^{*}$, $m\geq 0$ and $r_{-m} = r_{m}^{*}$, i.e., the autocorrelation of $\hat{\mathbf{c}}$.

\begin{figure}[tb]
\centering
\includegraphics[width=8.6cm]{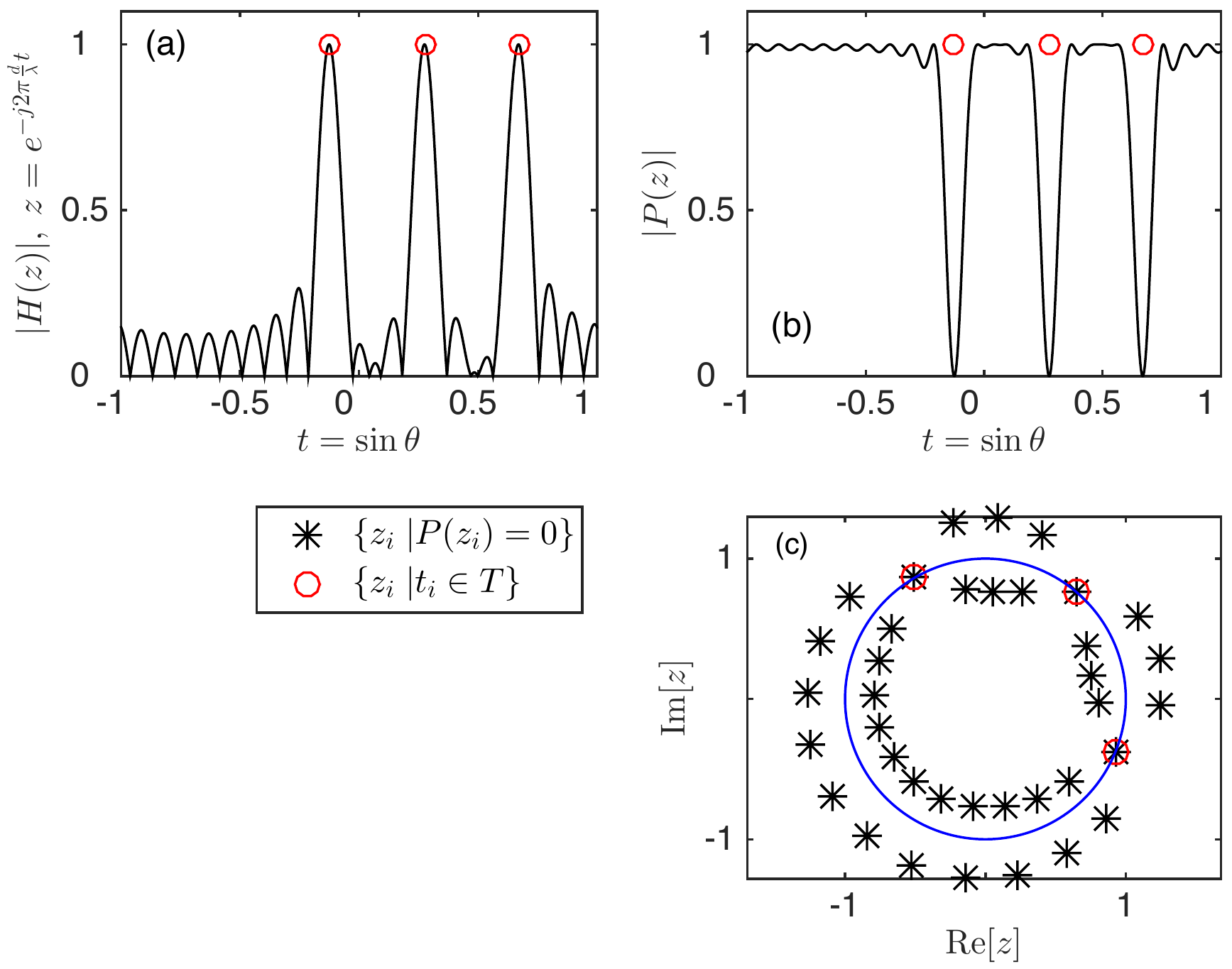}
\caption{(Color online) Support detection through the dual polynomial. A ULA is used with $M=21$ sensors and $d/ \lambda = 1/2$ to localize three sources with support set $T = [-0.126,0.275,0.67]$. (a) The dual polynomial $\lvert H(z)\rvert$. (b) The nonnegative polynomial $P(z)$. (c) The support $T$ is estimated by the angle of the roots, $z_{i}$, of $P(z)$ for which $\lvert z_{i}\rvert=1$.}
\label{fig:Roots3Sources21ULA}
\end{figure}

Note that the polynomial of degree $2(M-1)$,
\begin{equation}
\begin{aligned}
&P_{+}(z) = z^{M-1}P(z)\\
& =(1-r_{0})z^{M-1} - \sum\limits_{m=-(M-1),\, m\neq 0}^{M-1} r_{m} z^{\left(m+M-1\right)},
\end{aligned}
\label{eq:RootPolyEquiv}
\end{equation}
\noindent which has only positive powers of the variable $z$, has the same roots as $P(z)$, besides the trivial root $z=0$. Thus, the support $\hat{T}$ of $\hat{x}$, i.e., the DOAs of the sources, is recovered by locating the roots of $P_{+}(z)$ on the unit circle (see Fig.~\ref{fig:Roots3Sources21ULA}),
\begin{equation}
\hat{T} = \left\lbrace t_{i}=\frac{\lambda}{2\pi d} \arg z_{i} \; | \; P_{+}(z_{i})=0, \lvert z_{i}\rvert=1\right\rbrace. 
\label{eq:SupportRecovery}
\end{equation}

\subsection{Reconstruction of the primal solution $x$}

Once the support is recovered by locating the roots of the polynomial in Eq.~\eqref{eq:RootPolyEquiv} that lie on the unit circle~\eqref{eq:SupportRecovery}, the source amplitudes (the complex weights in Eq.~\eqref{eq:SuperpositionOfSpikes}) are recovered from,
\begin{equation}
\hat{\mathbf{x}}_{\text{CS}_{\text{dual}}} = \mathbf{A}_{T}^{+}\mathbf{y},
\label{eq:AmplitudeEstimation}
\end{equation}
where $^{+}$ denotes the pseudoinverse of $\mathbf{A}_{T}$ with columns $\mathbf{a}(t_{i})=e^{j2\pi(d/\lambda)\left[ 0,\cdots, M-1\right]^{T}t_{i}}$ for $t_{i}\in \hat{T}$.

\begin{figure}[tb]
\centering
\includegraphics[width=8.6cm]{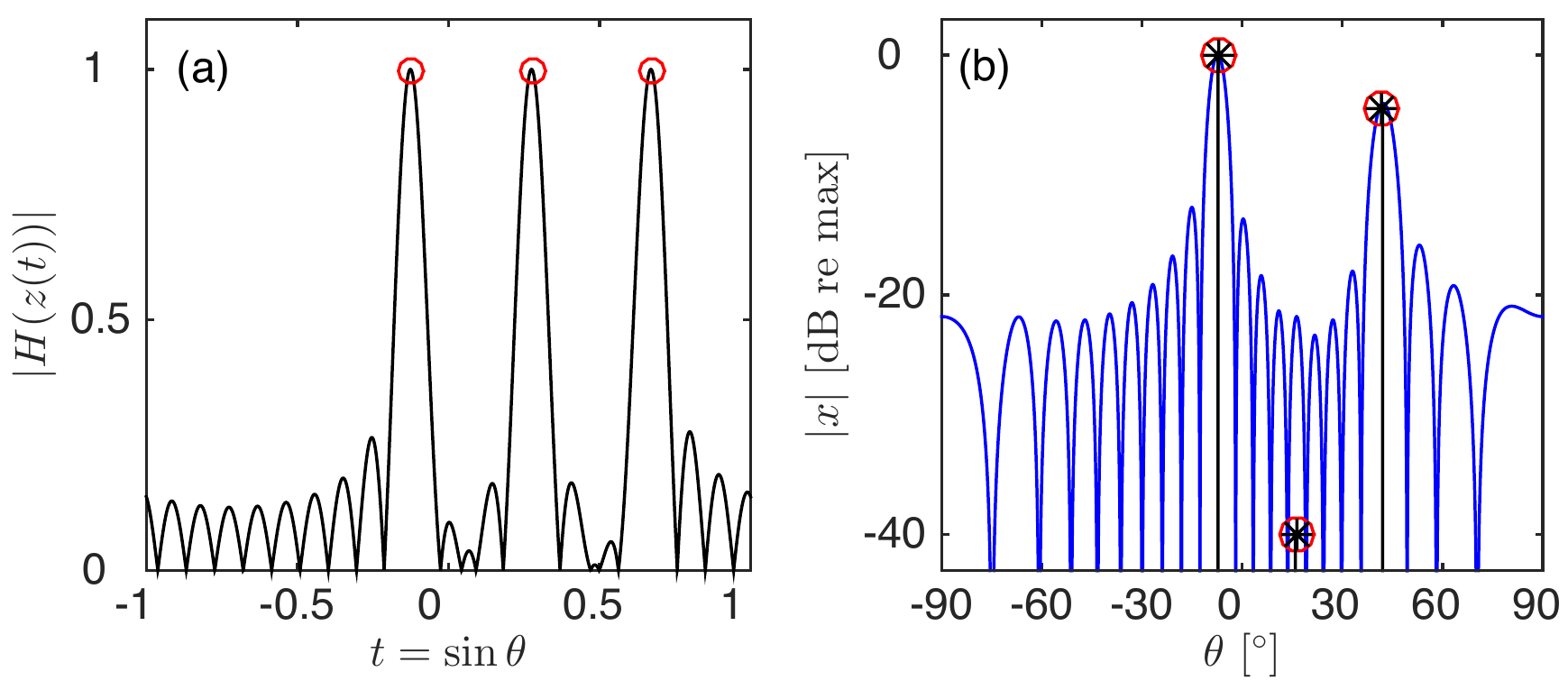}
\caption{(Color online) Grid-free sparse reconstruction. A standard ULA is used with $M=21$ sensors and $d/ \lambda = 1/2$ to localize three sources (o) at $\theta = [-7.2385^{\circ}, 15.962^{\circ}, 42.0671^{\circ}]$ with amplitudes $\lvert x \rvert = [1, 0.01, 0.6]$. (a) The dual polynomial. (b) Reconstruction with grid-free CS (*) and CBF.}
\label{fig:Solution3Sources21ULA}
\end{figure}

Figure~\ref{fig:Solution3Sources21ULA} shows the DOA estimation with grid-free CS following the procedure described in this section (see App.~\ref{sec:MatlabCode} for a Matlab implementation). The dual polynomial attains unit amplitude, $\lvert H\left(z\right)\rvert = 1$, at the support of the solution, i.e., the DOAs of the existing sources; see Fig.~\ref{fig:Solution3Sources21ULA}(a). Figure~\ref{fig:Solution3Sources21ULA}(b) compares the grid-free CS \eqref{eq:AmplitudeEstimation} and the CBF \eqref{eq:CBF} reconstruction in DOA estimation. The grid-free CS offers very accurate localization, while CBF is characterized by low resolution. Moreover, CBF fails to detect the weak source at $15.962^{\circ}$ since it is totally masked by the sidelobes.

\section{Maximum resolvable DOAs}

The maximum number of resolvable DOAs with grid-free CS is determined by the maximum number of roots of $P_{+}(z)$ in~\eqref{eq:RootPolyEquiv} which can be on the unit circle, $\lvert z\rvert=1$. Since the coefficients of the polynomial $P_{+}(z)$ are conjugate symmetric around the term $z^{M-1}$, the roots appear in pairs at the same angular direction $t_{l}$, one inside the unit circle, $z_{in} =r_{l}e^{-j2\pi\left(d/\lambda\right)t_{l}}$ at radius $r_{l}<1$, and the other outside of the unit circle $z_{out} =(1/r_{l})e^{-j2\pi\left(d/\lambda\right)t_{l}} = 1/(z_{in})^{H}$. This implies that the roots on the unit circle have double multiplicity. The polynomial $P_{+}(z)$ has in total $2(M-1)$ roots, as determined by its degree. Hence, there are at most $M-1$  (double) roots on the unit circle. 

The necessary condition for the dual polynomial~\eqref{eq:DualPolyAmplitude} to satisfy the condition $\lvert H(z) \rvert <1$ for some $t\in\mathbb{T}$, thus avoid the non-informative case of a constant dual polynomial, is that the number of sources should not exceed\cite{Fuchs:2005,GrandaSuperRes2014},
\begin{equation}
K_{max} = \left\lfloor\frac{M-1}{2}\right\rfloor,
\label{eq:MaxNumSources}
\end{equation}
\noindent where $\lfloor\cdot\rfloor$ is the largest integer not greater than the argument. In other words, at least half of the (paired) $M-1$ roots should lie off the unit circle alternating with the roots on the unit circle leading to the bound~\eqref{eq:MaxNumSources}.

For positive source amplitudes, $x_{i}\in\mathbb{R}_{+}$, the condition~\eqref{eq:MaxNumSources} is sufficient and no separation condition is required for the resolvable sources\cite{Fuchs:2005}. However, for complex amplitudes, $x_{i}\in\mathbb{C}$, the sources are resolved uniquely only if the corresponding DOAs are separated by at least \cite{GrandaSuperRes2014,TangCSoffGrid013},
\begin{equation}
\underset{t_{i},t_{j}\in T}{\min}\, \lvert t_{i}-t_{j} \rvert = \frac{\lambda}{M d},
\label{eq:MinSeparationCondition}
\end{equation}
\noindent where $\lvert t_{i}-t_{j} \rvert$ is a wrap-around distance meaning that we identify the points $-1$, $1$, in $\mathbb{T}=[-1,1]$.

The minimum separation condition~\eqref{eq:MinSeparationCondition} is a consequence of the coherence of the sensing process which is related to the beampattern; see Sec. IV.D. in Ref.~\onlinecite{XenakiCS:2014}. To guarantee a well-posed sparse signal reconstruction, it is required that the columns of the inverse Fourier operator $\mathcal{F}_{M}$, the steering vectors~\eqref{eq:SteeringVector}, are sufficiently uncorrelated. The continuous formulation~\eqref{eq:MeasurementVector} implies that adjacent steering vectors are in arbitrarily close directions, hence fully coherent. However, the requirement~\eqref{eq:MinSeparationCondition} inhibits closely spaced (i.e., highly correlated) steering vectors, hence prevents the sparse reconstruction problem from being too ill-posed due to coherence.

\begin{figure}[tb]
\centering
\includegraphics[width=8.6cm]{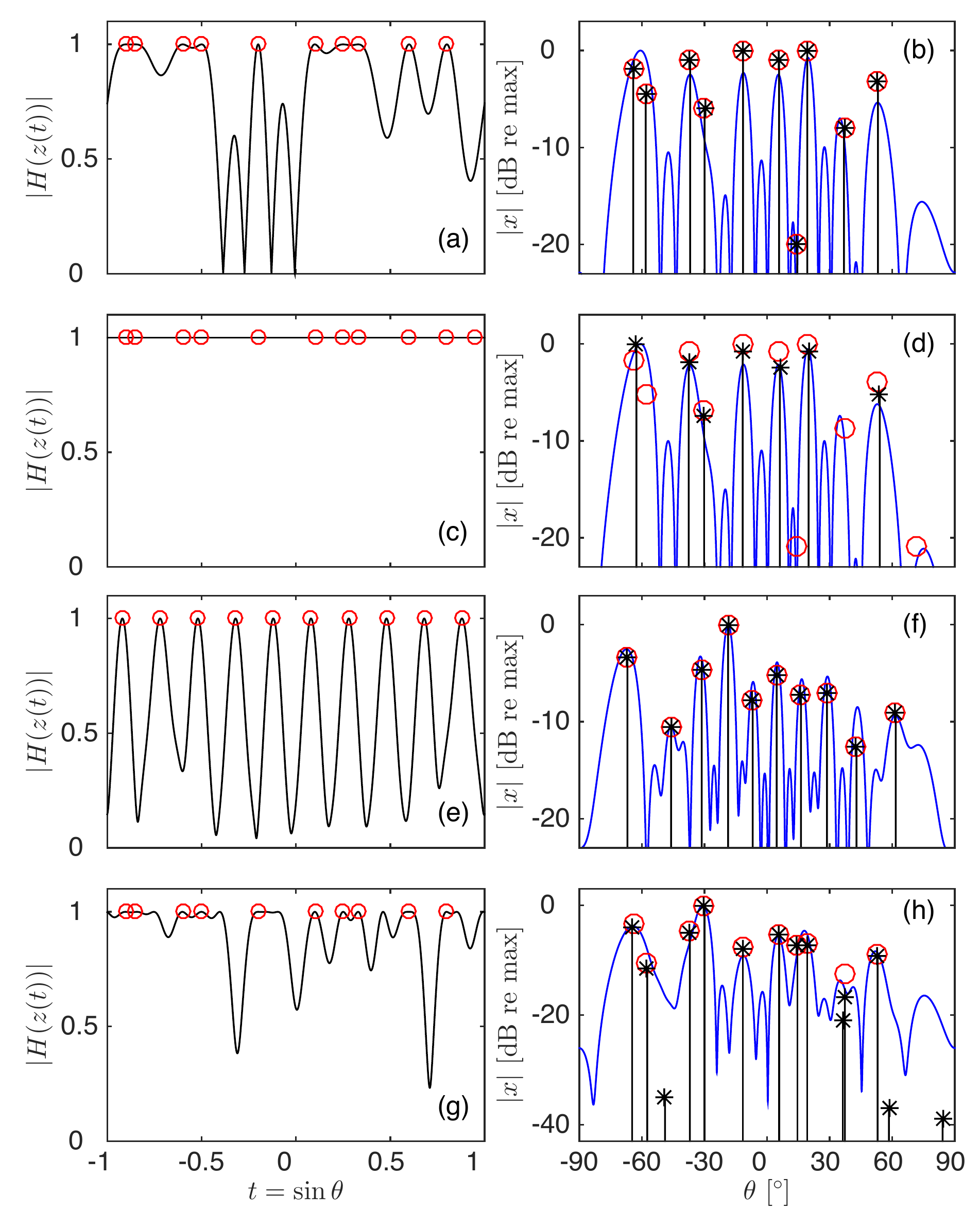}
\caption{(Color online) Grid-free sparse reconstruction. A ULA is used with $M=21$ sensors and $d/ \lambda = 1/2$ to localize the possible maximum number of sources (o), $\lfloor (M-1)/2 \rfloor= 10$. (a) The dual polynomial and (b) reconstruction with grid-free CS (*) and CBF for sources with positive amplitudes, $x_{10,\mathbb{R}} = [0.8, 0.6, 0.9, 0.5, 1, 0.9, 0.1, 1, 0.4, 0.7]$. (c) The dual polynomial and (d) reconstruction for 11 sources with positive amplitudes, $x_{11,\mathbb{R}} = [x_{10,\mathbb{R}}, 0.1]$. (e) The dual polynomial and (f) reconstruction for sources with complex amplitudes, $x_{10,\mathbb{C}} = x_{10,\mathbb{R}} +i[-1.6,0.5,-1.3,-2.6,0.4,-1.2,-1.2,-0.6,-0.5,0.6]$, separated by the condition~\eqref{eq:MinSeparationCondition}. (g) The dual polynomial and (h) reconstruction for sources with complex amplitudes, $x_{10,\mathbb{C}}$, but locations violating the condition~\eqref{eq:MinSeparationCondition}.}
\label{fig:Solution7Sources21ULA}
\end{figure}

Figure~\ref{fig:Solution7Sources21ULA} shows the reconstruction for the maximum number of sources possible. For positive source amplitudes, $x_{i} \in \mathbb{R}_{+}$, the bound~\eqref{eq:MaxNumSources} suffices to ensure a unique solution. Grid-free CS achieves super-resolution even for DOAs in general position; see Figs.~\ref{fig:Solution7Sources21ULA}(a)--(b). Inserting an additional source at 71.81$^{\circ}$, thus exceeding the maximum number of resolvable sources~\eqref{eq:MaxNumSources}, results in a non-informative dual polynomial, $\lvert H\left(z\right)\lvert \approx 1$, for all $t\in\mathbb{T}$, Fig.~\ref{fig:Solution7Sources21ULA}(c), and inaccurate reconstruction where only 7 out of the 11 sources are resolved, Fig.~\ref{fig:Solution7Sources21ULA}(d). For complex source amplitudes, $x_{i} \in \mathbb{C}$, an additional constraint~\eqref{eq:MinSeparationCondition} on the minimum separation of DOAs is required along with the bound on the number of sources~\eqref{eq:MaxNumSources} to ensure a unique solution, Figs.~\ref{fig:Solution7Sources21ULA}(e)--(f). Violating the minimum separation condition, the CS DOA estimation becomes extremely ill-posed due to the coherence of the underlying steering vectors resulting in inaccurate reconstruction characterized by the presence of spurious sources, Figs.~\ref{fig:Solution7Sources21ULA}(g)--(h).

\section{\label{sec:RandomArrays}Non-uniform arrays}

The method is also applicable to non-uniform arrays, constructed by randomly choosing sensors from a standard ULA configuration, by adding an additional constraint in the optimization problem~\eqref{eq:DualProblemSP}\cite{TangCSoffGrid013}. The additional constraint ensures that coefficients of the dual polynomial corresponding to inactive sensors on the ULA, $c_{m_{\text{null}}}$, are annihilated.

The dual problem in a semidefinite programming formulation~\eqref{eq:DualProblemSP} is augmented with an additional constraint and takes the form,
\begin{equation}
\begin{aligned}
\underset{\mathbf{c},\mathbf{Q}}{\max}\;\mathrm{Re} \left(\mathbf{c}^{H}\mathbf{y}\right) \; \text{subject to} \; \begin{bmatrix} \mathbf{Q}_{M\times M} & \mathbf{c}_{M\times 1}\\
\mathbf{c}^{H}_{1\times M} & 1\end{bmatrix} \succeq 0,&\\
\sum\limits_{i=1}^{M-j}\!\! \mathbf{Q}_{i,i+j} \!=\! \left\{ \begin{array}{rl}
 \!1, &j=0 \\
 \!0, &j=1,\cdots,M-1
       \end{array} \right.\!\!, \: c_{m_{\text{null}}} = 0. &
\end{aligned}
\label{eq:DualProblemSPRandom}
\end{equation}

\begin{figure}[tb]
\centering
\includegraphics[width=8.6cm]{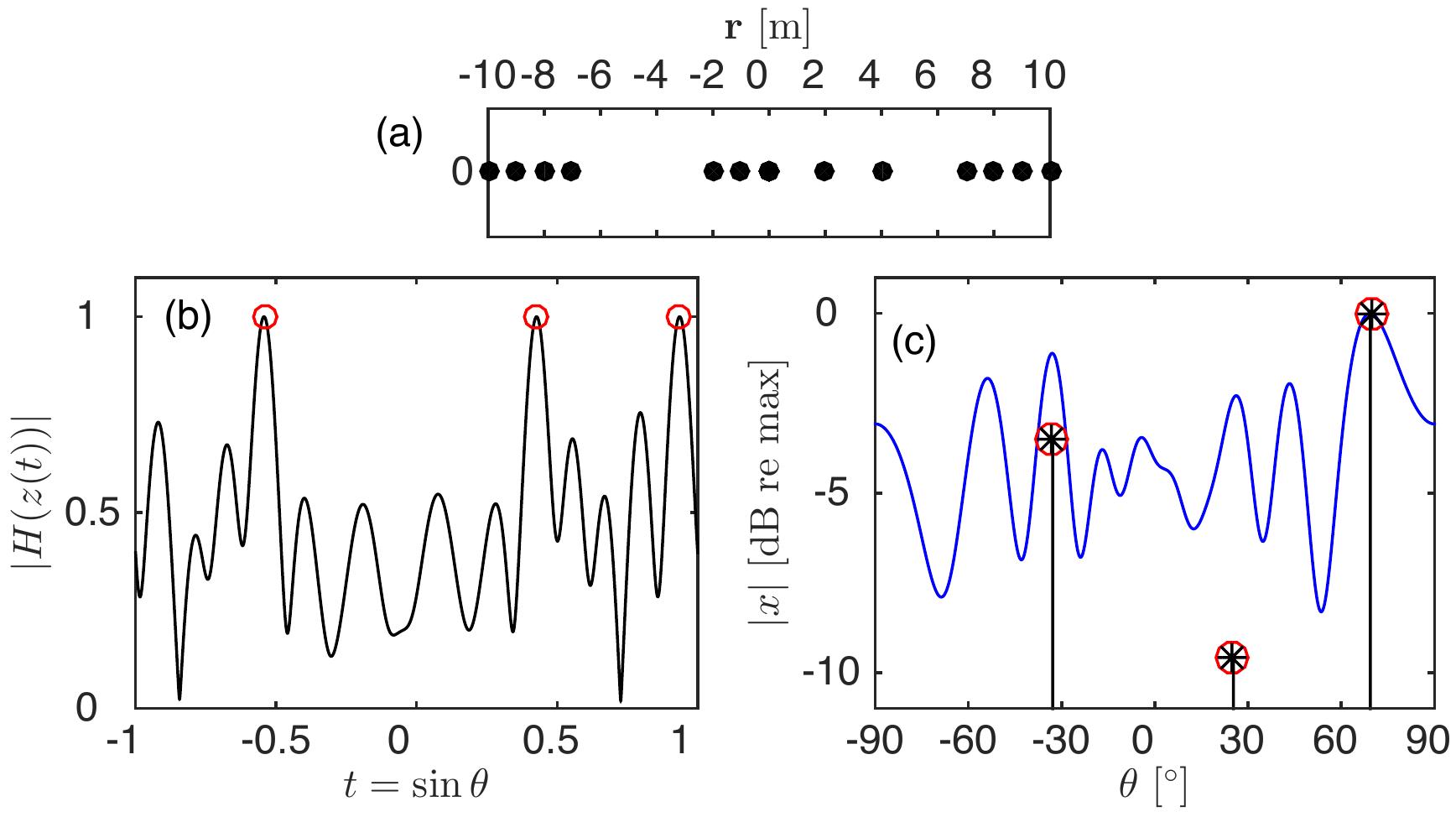}
\caption{(Color online) Grid-free sparse reconstruction. (a) A random array constructed by randomly selecting $M=13$ sensors out of a standard ULA with $21$ sensors and $d/ \lambda = 1/2$. The sources (o) are at $\theta = [-32.8881^{\circ}, 25.2773^{\circ}, 69.3903^{\circ}]$ with amplitudes $\lvert x \rvert = [0.67, 0.33, 1]$. (b) The dual polynomial. (c) Reconstruction with grid-free CS (*) and CBF.}
\label{fig:Solution3Sources13Random}
\end{figure}

Figure~\ref{fig:Solution3Sources13Random} shows the DOA estimation with grid-free CS and compares it with the CBF reconstruction in the case of a random array. Even though CBF performance degrades significantly due to the increased sidelobe levels introduced by the random array and the strong source towards endfire, CS still offers exact reconstruction.

\section{Grid-free reconstruction with noise}

The problem of grid-free DOA estimation with CS extends to noisy measurements making the framework useful for practical applications. Assuming that the measurements~\eqref{eq:NoisyDataContinuous} are contaminated with additive noise $\mathbf{n}\in \mathbb{C}^{M}$, such that $\lVert\mathbf{n}\rVert_{2}\leq \epsilon$, the atomic norm minimization problem~\eqref{eq:PrimalProblem} is reformulated as\cite{GrandaNoisyData2013},
\begin{equation}
\underset{x}{\min}\lVert x\rVert_{\mathcal{A}} \; \text{subject to} \;  \lVert \mathbf{y} - \mathcal{F}_{M}x\rVert_{2}\leq\epsilon. 
\label{eq:PrimalProblemNoise}
\end{equation}
To solve the infinite dimensional primal problem~\eqref{eq:PrimalProblemNoise} we formulate the equivalent dual problem (see Appendix~\ref{sec:AppDualNoise}),
\begin{equation}
\underset{\mathbf{c}}{\max}\;\mathrm{Re}\left(\mathbf{c}^{H}\mathbf{y}\right) -\epsilon\lVert\mathbf{c}\rVert_{2} \; \text{subject to} \;  \lVert\mathcal{F}^{H}_{M}\mathbf{c}\rVert_{\infty}\leq 1,
\label{eq:DualProblemNoise}
\end{equation}
\noindent and we replace the infinite-dimensional constraints with finite matrix inequalities,  
\begin{equation}
\begin{aligned}
\underset{\mathbf{c},\mathbf{Q}}{\max}\:\mathrm{Re}\!\left(\mathbf{c}^{H}\mathbf{y}\right) \!-\! \epsilon\lVert\mathbf{c}\rVert_{2} \; \text{sub. to} \begin{bmatrix} \!\mathbf{Q}_{M\times M} & \!\mathbf{c}_{M\times 1}\!\\
\!\mathbf{c}^{H}_{1\times M} & \!1 \!\end{bmatrix} \!\succeq \!0,&\\
\sum\limits_{i=1}^{M-j} \mathbf{Q}_{i,i+j} = \left\{ \begin{array}{rl}
 1, &j=0 \\
 0, &j=1,\cdots,M-1.
       \end{array} \right.&
\end{aligned}
\label{eq:DualProblemSPNoise}
\end{equation}

The problem~\eqref{eq:DualProblemSPNoise} is a convex optimization problem which can be solved efficiently with semidefinite programming\cite{CVX} to obtain an estimate for the coefficients, $\mathbf{c}\in \mathbb{C}^{M}$, of the dual polynomial. The support of the solution, i.e., the DOAs of the existing sources is found by locating the points where the dual polynomial has unit amplitude following the methodology in Sec~\ref{sec:SupportDetection}. Once the support is recovered the source amplitudes are estimated by solving a discrete overdetermined problem~\eqref{eq:AmplitudeEstimation}.

\begin{figure}[tb]
\centering
\includegraphics[width=8.6cm]{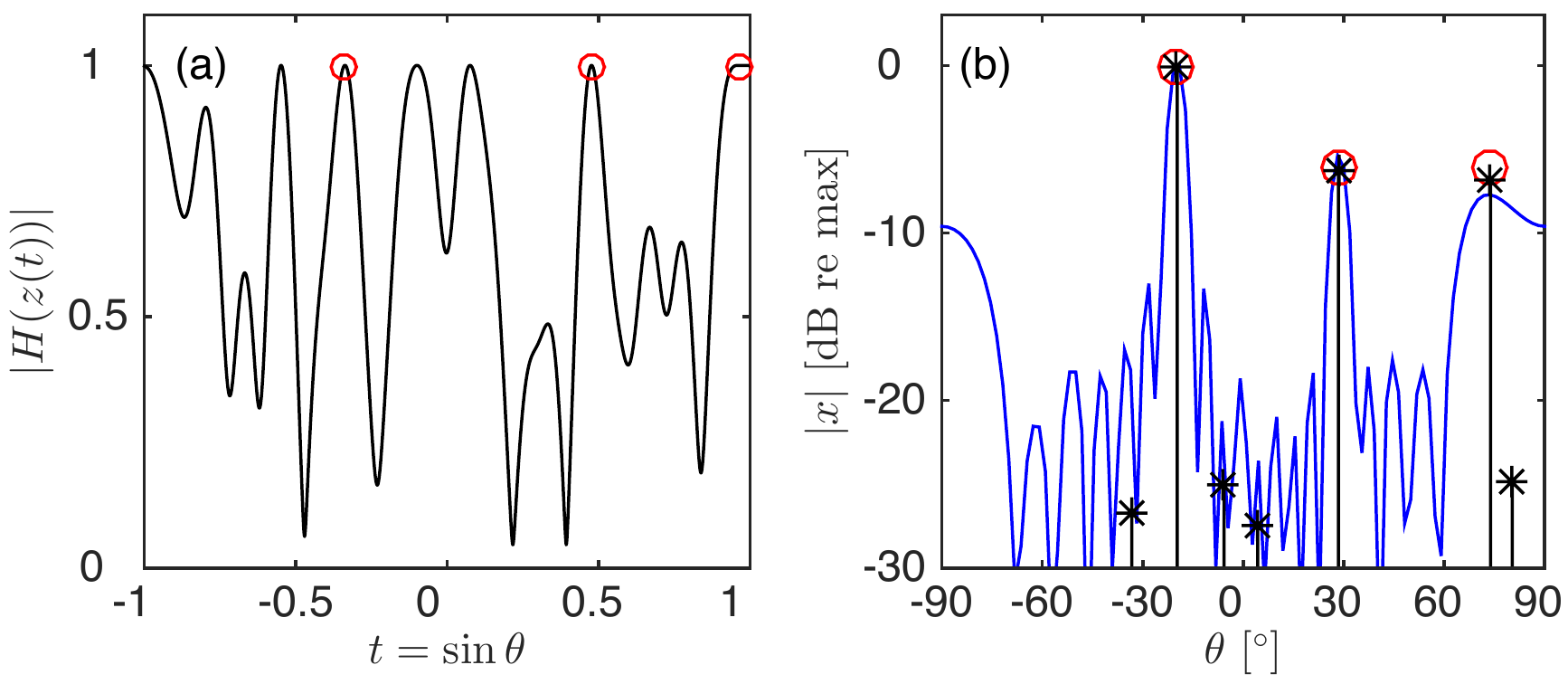}
\caption{(Color online) Grid-free sparse reconstruction. A ULA is used with $M=21$ sensors and $d/ \lambda = 1/2$ to localize three sources (o) at $\theta = [-19.6942^{\circ}, 28.3594^{\circ}, 73.9457^{\circ}]$ with amplitudes $\lvert x \rvert = [0.6, 0.3, 0.3]$. (a) The dual polynomial. (b) Reconstruction with grid-free CS (*) and CBF. The SNR is $20$ dB.}
\label{fig:SolutionNoise20SNR3sources21ULA}
\end{figure}

Figure~\ref{fig:SolutionNoise20SNR3sources21ULA} shows the DOA estimation for three sources with grid-free CS when the array measurements are contaminated with additive noise~\eqref{eq:NoisyDataContinuous} such that SNR=$20$~dB. Grid-free CS improves significantly the resolution in the reconstruction compared to CBF, even though some weak spurious sources appear as artifacts due to the noise in the measurements.

\section{\label{sec:RootMethods}DOA estimation with polynomial rooting}

Polynomial rooting can increase performance and achieve super-resolution in several DOA estimation methods, such as the minimum variance distortionless response (MVDR) beamformer, the multiple signal classification (MUSIC) method and the minimum-norm method. All these methods involve the estimation or the eigendecomposition of the cross-spectral matrix both in their spectral and root version.

The cross-spectral matrix estimated from $L$ snapshots (i.e., observations of $\mathbf{y}$ at a particular frequency) is defined as,
\begin{equation}
\hat{\mathbf{C}}_{\mathbf{y}}= \frac{1}{L}\sum\limits_{l=1}^{L}\mathbf{y}_{l}\mathbf{y}_{l}^{H}.
\label{eq:CrossSpectralMatrix}
\end{equation}

The eigendecomposition of the cross-spectral matrix separates the signal and the noise subspaces, 
\begin{equation}
\hat{\mathbf{C}}_{\mathbf{y}} = \hat{\mathbf{U}}_{s}\hat{\mathbf{\Lambda}}_{s}\hat{\mathbf{U}}_{s}^{H} +\hat{\mathbf{U}}_{n}\hat{\mathbf{\Lambda}}_{n}\hat{\mathbf{U}}_{n}^{H},
\label{eq:CSMeigendecomposition}
\end{equation}
\noindent where $\hat{\mathbf{U}}_{s}$ comprises the signal eigenvectors, which correspond to the largest eigenvalues $\hat{\mathbf{\Lambda}}_{s}$, and $\hat{\mathbf{U}}_{n}$ comprises the noise eigenvectors. The signal eigenvectors are in the same subspace as the steering vectors~\eqref{eq:SteeringVector}, while the noise eigenvectors are orthogonal to the subspace of the steering vectors, thus $\mathbf{a}(\theta)^{H}\hat{\mathbf{U}}_{n} = \mathbf{0}$. 

\subsection{Spectral version of DOA estimation methods}

MVDR\cite{Capon:1969} aims to minimize the output power of the beamformer under the constraint that the signal from the look direction remains undistorted. The MVDR beamformer power spectrum is,
\begin{equation}
P_{\text{MVDR}}(\theta) =  \frac{1}{\mathbf{a}(\theta)^{H}\hat{\mathbf{C}}_{\mathbf{y}}^{-1}\mathbf{a}(\theta)}.
\label{eq:Pmvdr}
\end{equation}

MUSIC\cite{SchmidtMUSIC:1986} uses the orthogonality between the signal and the noise subspace to locate the maxima in the spectrum, 
\begin{equation}
P_{\text{MUSIC}}(\theta) = \frac{1}{\mathbf{a}(\theta)^{H}\hat{\mathbf{U}}_{n}\hat{\mathbf{U}}_{n}^{H}\mathbf{a}(\theta)}.
\label{eq:Pmusic}
\end{equation}

The minimum-norm is also an eigendecomposition based method but, unlike MUSIC which utilizes all noise eigenvectors, it uses a single vector, $\mathbf{v} = [v_{0}, \cdots, v_{M-1}]^{T}$, which resides in the noise subspace (compare with the dual vector $\hat{\mathbf{c}}$~\eqref{eq:DualConstraint} which resides in the signal subspace) such that,
\begin{equation}
\mathbf{a}(\theta_{i})^{H} \mathbf{v} = 0, \quad i= 1, \cdots, K,
\label{eq:MinNormVector}
\end{equation}
\noindent where $K$ is the number of sources.

All the noise subspace eigenvectors, i.e., the columns of $\hat{\mathbf{U}}_{n}$ have the property in Eq.~\eqref{eq:MinNormVector}. However, if the vector $\mathbf{v}$ is chosen as a linear combination of the noise subspace eigenvectors the algorithm tends to be more robust \cite{KumaresanMinNorm:1983,KumaresanZeros:1983, VanTreesBook}.

The minimum-norm method selects a vector, $\mathbf{v}$, in the noise subspace with minimum $\ell_{2}$-norm and unit first element, $v_{0} = 1$. The vector $\mathbf{v}$ can be constructed from the noise eigenvectors as\cite{KumaresanMinNorm:1983},
\begin{equation}
\mathbf{v} = \hat{\mathbf{U}}_{n}\mathbf{d}^{H}/\lVert\mathbf{d}\rVert_{2}^{2},
\label{eq:MinNormVectorNoiseSubspace}
\end{equation}
\noindent where the vector $\mathbf{d}$ is the first row of $\hat{\mathbf{U}}_{n}$. Equivalently, the vector $\mathbf{v}$ can be constructed from the signal eigenvectors as,
\begin{equation}
\mathbf{v} = \hat{\mathbf{U}}_{s}\frac{\mathbf{b}^{H}}{1-\lVert\mathbf{b}\rVert_{2}^{2}},
\label{eq:MinNormVectorSignalSubspace}
\end{equation}
\noindent where the vector $\mathbf{b}$ is the first row of $\hat{\mathbf{U}}_{s}$.

The minimum-norm spectrum is,
\begin{equation}
P_{\text{min-norm}}(\theta) = \frac{1}{\mathbf{a}(\theta)^{H}\mathbf{v}\mathbf{v}^{H}\mathbf{a}(\theta)}.
\label{eq:Pminnorm}
\end{equation}

\subsection{\label{sec:RootDOA}Root version of DOA estimation methods}

The root version of the DOA estimation methods is based on the fact that for ULAs the null spectrum has the form of the trigonometric polynomial in Eq.~\eqref{eq:NNTP} with $\omega = 2\pi(d/\lambda) \sin\theta$  (since $\sin\theta \in [-1,1]$, then for a standard ULA $\omega \in \left[ -\pi, \pi \right]$). Thus, evaluating the spectrum is equivalent to evaluating the roots of the polynomial on the unit circle \cite{Barabell:1983}.

More analytically, let $N(\theta)=\mathbf{a}(\theta)^{H} \mathbf{\Psi} \mathbf{a}(\theta)$ be the null spectrum, such that the spectrum is $S(\theta)=N(\theta)^{-1}$. For MVDR, $\mathbf{\Psi}= \hat{\mathbf{C}}_{\mathbf{y}}^{-1}$ (Ref.\onlinecite{VanTreesBook}, p.1147), for MUSIC, $\mathbf{\Psi}= \hat{\mathbf{U}}_{n}\hat{\mathbf{U}}_{n}^{H}$ (Ref.\onlinecite{VanTreesBook}, p.1159) and for the minimum-norm method, $\mathbf{\Psi}= \mathbf{v}\mathbf{v}^{H}$ (Ref.\onlinecite{VanTreesBook}, p.1163). Then,
\begin{equation}
\begin{aligned}
N(\theta) &= \sum\limits_{m=0}^{M-1} \sum\limits_{n=0}^{M-1} e^{-j2\pi m \frac{d}{\lambda}\sin\theta} \Psi_{mn} e^{-j2\pi n \frac{d}{\lambda}\sin\theta}\\
&= \sum\limits_{l=-(M-1)}^{M-1} \psi_{l} e^{-j2\pi l \frac{d}{\lambda}\sin\theta}\\
N(z) &= \sum\limits_{l=-(M-1)}^{M-1} \psi_{l} z^{-l},
\end{aligned}
\label{eq:Pol4RootMethods}
\end{equation}
\noindent where $\psi_{l} = \sum_{m-n=l} \Psi_{mn}$  is the sum of the elements of the Hermitian matrix $\mathbf{\Psi}$ along the $l$th diagonal and $z = e^{j2\pi (d/\lambda)\sin\theta}$.

The set of DOAs, $\hat{T}$, is estimated from the roots of the polynomial $N(z)$, or equivalently the polynomial $N_{+}(z) = z^{M-1}N(z)$, which lie on the unit circle, $z_{i} =  e^{j \arg(z_{i})}$ as,
\begin{equation}
\hat{T} = \left\lbrace \sin\theta_{i}=\frac{\lambda}{2\pi d} \arg z_{i} \; | \; N_{+}(z_{i})=0, \lvert z_{i}\rvert=1\right\rbrace. 
\label{eq:SupportRecoveryRootMethods}
\end{equation}
\noindent After the support is recovered, the amplitudes can be estimated through an overdetermined problem as in Eq.~\eqref{eq:AmplitudeEstimation}.

Even though the root forms of DOA estimation methods have, often, more robust performance than the corresponding spectral forms \cite{Rao:1989}, they require a regular array geometry to form a trigonometric polynomial and detect its roots behavior. To achieve a robust estimate of the cross-spectral matrix many snapshots are required, $L>M$, i.e., stationary sources. Furthermore, eigendecomposition based methods fail to discern coherent arrivals. Forward/backward smoothing techniques\cite{Pillai:1989, RaoSpatialSmoothing:1990} can be employed to mitigate this problem and make eigendecomposition based methods suitable for identification of coherent sources as well, but they still require a regular array geometry and an increased number of sensors.

\section{Experimental results}

The high-resolution capabilities of sparse signal reconstruction methods, i.e., CS for DOA estimation, and the robustness of grid-free sparse reconstruction even under noisy conditions and with random array configurations are demonstrated on ocean acoustic measurements. The interest is on single-snapshot reconstruction for source tracking and the results are compared with CBF.  

The data set is from the long range acoustic communications (LRAC) experiment\cite{FORAdata:2011} recorded from 10:00-10:30 UTC on $16$ September 2010 in the NE Pacific and is the same as in Ref.\onlinecite{XenakiCS:2014} to allow comparison of the results. The data are from a horizontal uniform linear array towed at $3.5$ knots at $200$~m depth. The array has $M=64$ sensors, with intersensor spacing $d=3$~m. The data were acquired with a sampling frequency of $2000$~Hz and the record is divided in $4$~s non-overlapping snapshots. Each snapshot is Fourier transformed with $2^{13}$ samples.  

\begin{figure}[tb]
\centering
\includegraphics[width=8.6cm]{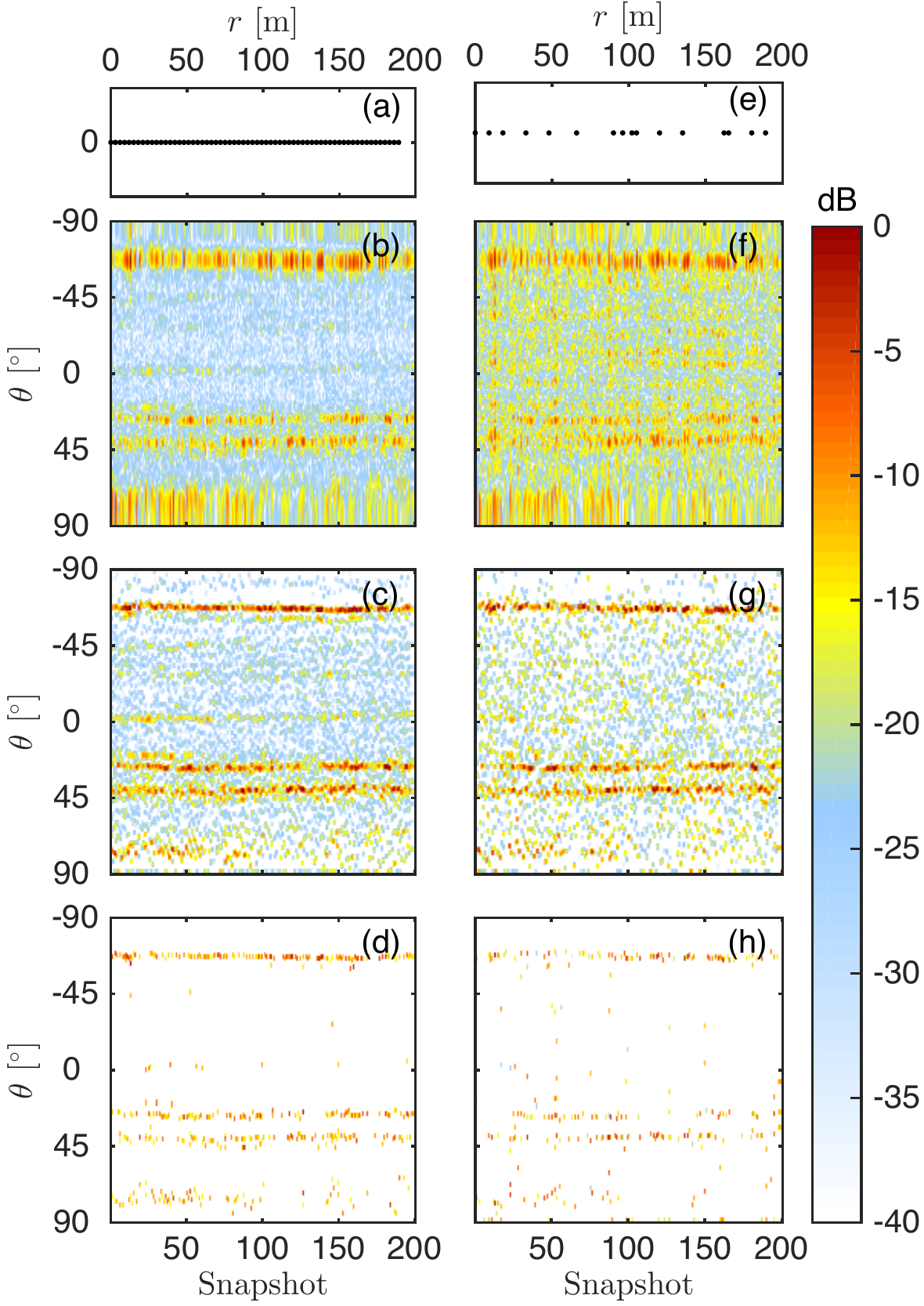}
\caption{(Color online) Data from LRAC. (a) Uniform array with $M=64$ sensors and the corresponding (b) CBF, (c) CS on a discrete grid, [$-90^{\circ}$:$1^{\circ}$:$90^{\circ}$], and (d) grid-free CS reconstruction. (e) Non-uniform array with $M=16$ sensors and the corresponding (f) CBF, (g) CS on a discrete grid and (h) grid-free CS reconstruction.}
\label{fig:DualLRACData}
\end{figure}

The data are post-processed with CBF and CS on a discrete DOA grid [$-90^{\circ}$:$1^{\circ}$:$90^{\circ}$] as well as grid-free CS at frequency $f=125$~Hz ($d/\lambda=1/4$). To facilitate the comparison of the results, the grid-free CS reconstruction is also presented on the grid [$-90^{\circ}$:$1^{\circ}$:$90^{\circ}$] by rounding the estimated DOAs to the closest integer angle and using the maximum power within each bin. The results are depicted in Fig.~\ref{fig:DualLRACData} both with all $M=64$ sensors active, Figs.~\ref{fig:DualLRACData}(a)--(d) and by retaining only $M=16$ sensors active in a non-uniform configuration, Figs.~\ref{fig:DualLRACData}(e)--(h). Both array configurations, Figs.~\ref{fig:DualLRACData}(a) and~\ref{fig:DualLRACData}(e), have the same aperture thus the same resolution.

The CBF map~\eqref{eq:CBF} in Fig.~\ref{fig:DualLRACData}(b) indicates the presence of three stationary sources at around $45^{\circ}$, $30^{\circ}$ and $-65^{\circ}$. The two arrivals at $45^{\circ}$ and $30^{\circ}$ are attributed to distant transiting ships, even though a record of ships in the area was not kept. The broad arrival at $-65^{\circ}$ is from the towship R/V Melville. The CBF map suffers from low resolution and artifacts due to sidelobes and noise. The CS reconstruction~\eqref{eq:CS_solution_noisy} ($\epsilon$=3.5, Fig.~\ref{fig:DualLRACData}(c)) results in improved resolution in the localization of the three sources by promoting sparsity and significant reduction of artifacts in the map. The grid-free CS solution~\eqref{eq:AmplitudeEstimation}, Fig.~\ref{fig:DualLRACData}(d), provides high resolution and further artifact reduction due to polynomial rooting.

Retaining only $1/4$ of the sensors on the array in a non-uniform configuration degrades the resolution of CBF due to increased sidelobe levels, Fig.~\ref{fig:DualLRACData}(f). However, both CS on a discrete DOA grid, Fig.\ref{fig:DualLRACData}(g), and grid-free CS, Fig.\ref{fig:DualLRACData}(h), provide high-resolution DOA estimation without a significant reconstruction degradation.

\begin{figure}[tb]
\centering
\includegraphics[width=8.6cm]{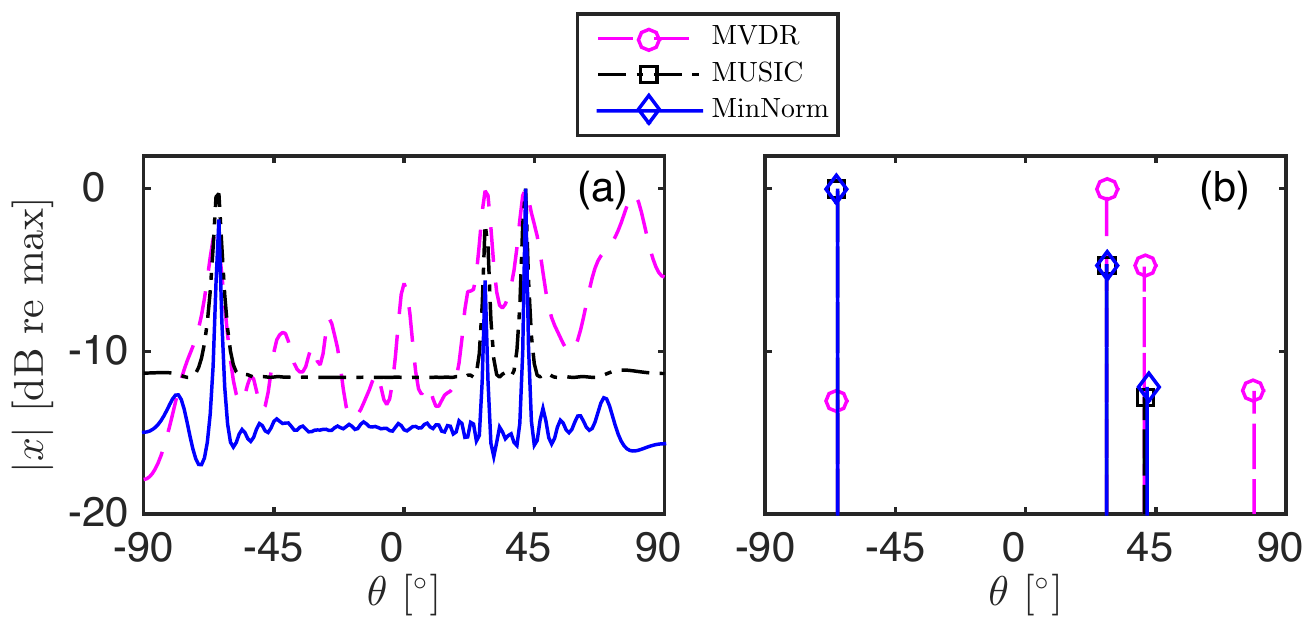}
\caption{(Color online) Data from LRAC, combining the 200 snapshots to estimate the cross-spectral matrix and processing with MVDR, MUSIC and the minimum-norm method. (a) Spectral version and (b) root version. The ULA with $M=64$ sensors and $d/\lambda=1/4$ is used.}
\label{fig:DualLRACRootMethods}
\end{figure}

The single-snapshot processing, Fig.~\ref{fig:DualLRACData}, indicates that the sources are adequately stationary. Therefore, the 200 snapshots can be combined to estimate the cross-spectral matrix \eqref{eq:CrossSpectralMatrix} and employ cross-spectral methods for DOA estimation. Figure~\ref{fig:DualLRACRootMethods}(a) compares the power spectra of MVDR \eqref{eq:Pmvdr}, MUSIC \eqref{eq:Pmusic} and the minimum-norm method \eqref{eq:Pminnorm} and Fig.~\ref{fig:DualLRACRootMethods}(b) the corresponding root versions.

The root versions of cross-spectral methods, especially the root MUSIC and the root minimum-norm method, provide improved resolution compared to the corresponding spectral forms. However, the root cross-spectral methods require both many snapshots (i.e., stationary sources) for a robust estimate of the cross-spectral matrix and uniform arrays. Grid-free CS does not have these limitations.

\section{Conclusion}

DOA estimation with sensor arrays is a sparse signal reconstruction problem which can be solved with compressive sensing (CS). Discretization of the problem involves a compromise between the quality of reconstruction and the computational complexity, especially for high-dimensional problems. Grid-free CS assures that the sparsity promoting optimization problem in CS can be solved in the dual domain with semidefinite programming even when the unknowns are infinitely many. Grid-free CS achieves high-resolution DOA estimation through the polynomial rooting method.

In contrast to established DOA estimation methods, CS provides high-resolution acoustic imaging even with non-uniform array configurations and robust performance under noisy measurements and single-snapshot data. Finally, the grid-free CS has the same performance both with coherent and incoherent, stationary or moving sources while other DOA estimation methods based on polynomial rooting fail to discern coherent arrivals and have degraded resolution for moving sources as they require many snapshots.

\section*{ACKNOWLEDGMENTS}

This work was supported by the Office of Naval Research, under Grant No. N00014-11-1-0320.

\appendix
\section{\label{sec:AppOptimizationProblems}Convex optimization problems}

This section summarizes the basic notions and formulations encountered in convex optimization problems, as presented analytically in Ref.~\onlinecite{BoydBook}.

\subsection{Primal problem}

A generic optimization problem has the form,
\begin{equation}
\begin{aligned}
& \underset{\mathbf{x}}{\min} \; f_{0}(\mathbf{x})\\
& \text{subject to} \; & f_{i}(\mathbf{x})\leq 0, \; i=1,\cdots, m\\
& & h_{j}(\mathbf{x}) =0, \; j=1,\cdots, q,
\end{aligned}
\label{eq:ConvexPrimal}
\end{equation}
where $\mathbf{x}\in\mathbb{C}^{N}$ is the optimization variable, the function $f_{0}:\mathbb{C}^{N}\rightarrow \mathbb{R}$ is the objective (or cost) function, the functions $f_{i}:\mathbb{C}^{N}\rightarrow \mathbb{R}$ are the inequality constraint functions and the functions $h_{j}:\mathbb{C}^{N}\rightarrow \mathbb{C}$ are the equality constraint functions. The optimization problem~\eqref{eq:ConvexPrimal} is convex when $f_{0},\cdots, f_{m}$ are convex functions and $h_{1},\cdots, h_{q}$ are affine (linear) functions.

The set of points for which the objective and all constraint functions in Eq.~\eqref{eq:ConvexPrimal} are defined is called the domain of the optimization problem,
\begin{equation}
\mathcal{D} = \bigcap\limits_{i=0}^{m} \text{dom} f_{i} \cap \bigcap\limits_{j=1}^{q} \text{dom} h_{j}.
\label{eq:Domain}
\end{equation}
A point $\tilde{\mathbf{x}}\in\mathcal{D}$ is called feasible if it satisfies the constraints in Eq.~\eqref{eq:ConvexPrimal}.

The optimal value $p^{*}$ of the optimization problem~\eqref{eq:ConvexPrimal}, achieved at the optimal variable $\mathbf{x}^{*}$, is,
\begin{equation}
\begin{aligned}
p^{*} &= \inf\left\lbrace f_{0}(\mathbf{x}) \;|\; f_{i}(\mathbf{x})\leq 0, \; h_{j}(\mathbf{x}) =0\right\rbrace\\
&= \left\lbrace f_{0}(\mathbf{x}^{*}) \;|\; f_{i}(\mathbf{x}^{*})\leq 0, \; h_{j}(\mathbf{x}^{*}) =0\right\rbrace,
\end{aligned}
\label{eq:PrimalOptimum}
\end{equation}
\noindent for all $i=1,\cdots, m$ and $j=1,\cdots, q$.

\subsection{The Lagrangian}

The Lagrangian, $L$, of an optimization problem is obtained by augmenting the objective function with a weighted sum of the constraint functions. The Lagrangian of the generic optimization problem~\eqref{eq:ConvexPrimal} is ,
\begin{equation}
L(\mathbf{x},\bm{\lambda},\bm{\nu}) \!= \! f_{0}(\mathbf{x}) + \sum_{i=1}^{m}\lambda_{i} f_{i}(\mathbf{x})  + \mathrm{Re}\! \left[\sum_{j=1}^{q} \nu_{i} h_{i}(\mathbf{x})\right]\!,
\label{eq:Lagrangian}
\end{equation}
where $\lambda_{i}$ is the Lagrange multiplier associated with the $i$th inequality constraint, $f_{i}(\mathbf{x})\leq 0$, and $\nu_{j}$ is the Lagrange multiplier associated with the $j$th equality constraint, $h_{j}(\mathbf{x})=0$. The vectors $\bm{\lambda}\in\mathbb{R}^{m}$ and $\bm{\nu}\in \mathbb{C}^{q}$ are the dual variables of the problem~\eqref{eq:ConvexPrimal}.

\subsection{The dual function}

The dual function of the problem~\eqref{eq:ConvexPrimal} is the minimum value of the Lagrangian~\eqref{eq:Lagrangian} over $\mathbf{x}\in\mathcal{D}$ for $\bm{\lambda}\in\mathbb{R}^{m}$ and $\bm{\nu}\in \mathbb{C}^{q}$,
\begin{equation}
g(\bm{\lambda},\bm{\nu}) =\underset{\mathbf{x}\in\mathcal{D}}{\inf} L(\mathbf{x},\bm{\lambda},\bm{\nu}).
\label{eq:DualFunction}
\end{equation}
Since the dual function is the pointwise infinum of a family of affine functions of $(\bm{\lambda},\bm{\nu})$, it is concave, even when the problem~\eqref{eq:ConvexPrimal} is not convex.

The dual function~\eqref{eq:DualFunction} yields lower bounds on the optimal value $p^{*}$~\eqref{eq:PrimalOptimum} for any $\bm{\lambda}\succeq 0$ (where $\succeq$ represents componentwise inequality) and any $\bm{\nu}$,
\begin{equation}
g(\bm{\lambda},\bm{\nu}) \leq p^{*},
\label{eq:LowerBound}
\end{equation}
since $g(\bm{\lambda},\bm{\nu}) =\underset{\mathbf{x}\in\mathcal{D}}{\inf} L(\mathbf{x},\bm{\lambda},\bm{\nu})\leq L(\tilde{\mathbf{x}},\bm{\lambda},\bm{\nu})\leq f_{0}(\tilde{\mathbf{x}})$ for every feasible point $\tilde{\mathbf{x}}$.

\subsection{Dual problem}

The dual function~\eqref{eq:DualFunction} gives a lower bound on the optimal value $p^{*}$ of the optimization problem~\eqref{eq:ConvexPrimal}, which depends on the dual variables $(\bm{\lambda},\bm{\nu})$ with $\bm{\lambda}\succeq 0$; see Eq.~\eqref{eq:LowerBound}. The best lower bound, i.e. the lower bound with the greatest value, is obtained through the optimization problem, 
\begin{equation}
\underset{\bm{\lambda, \bm{\nu}}}{\max}\; g(\bm{\lambda},\bm{\nu}) \; \text{subject to} \; \bm{\lambda} \succeq 0,
\label{eq:ConvexDual}
\end{equation}
\noindent which is the dual problem to the optimization problem~\eqref{eq:ConvexPrimal}.
 
The dual problem~\eqref{eq:ConvexDual} is a convex optimization problem, since the objective function to be maximized is concave and the constraints are convex, irrespectively whether the primal problem~\eqref{eq:ConvexPrimal} is convex or not.

\subsection{Weak duality}

The optimal value $d^{*}$ of the dual problem~\eqref{eq:ConvexDual}, achieved at the dual optimal variables $(\bm{\lambda}^{*},\bm{\nu}^{*})$ is,
\begin{equation}
\begin{aligned}
d^{*} &= \sup\left\lbrace g(\bm{\lambda},\bm{\nu}) \;|\;  \bm{\lambda} \succeq 0\right\rbrace\\
&= \left\lbrace g(\bm{\lambda}^{*},\bm{\nu}^{*}) \;|\;  \bm{\lambda}^{*} \succeq 0\right\rbrace.
\end{aligned}
\label{eq:DualOptimum}
\end{equation}
\noindent The dual maximum $d^{*}$ is the best lower bound on the minimum of the primal problem~\eqref{eq:PrimalOptimum}, that can be obtained from the Lagrange dual function. The inequality,
\begin{equation}
d^{*}\leq p^{*},
\label{eq:WeakDuality}
\end{equation}
\noindent holds even if the primal problem~\eqref{eq:ConvexPrimal} is non-convex and is called weak duality.

The non-negative difference $p^{*}-d^{*}$ is called the duality gap for the optimization problem~\eqref{eq:ConvexPrimal}, since it gives the gap between the minimum of the primal problem and the maximum of the dual problem.

\subsection{Slater's condition and strong duality}

When the duality gap, $p^{*}-d^{*}$, is zero, strong duality holds characterized by the equality,  
\begin{equation}
d^{*}=p^{*}.
\label{eq:StrongDuality}
\end{equation}

Strong duality holds when the optimization problem~\eqref{eq:ConvexPrimal} is convex and there exists a strictly feasible point, i.e., the inequality constraints hold with strict inequalities. The constraint qualification which implies strong duality for convex problems is called Slater's condition,
\begin{equation}
\begin{aligned}
& f_{i}(\mathbf{x})<0, \; i = 1,\cdots,m,\\
& \mathbf{A}_{q\times N}\mathbf{x} = \mathbf{y}.
\end{aligned}
\label{eq:SlatersCondition}
\end{equation}
\noindent When the primal problem is convex and Slater's condition holds there exist a dual feasible $(\bm{\lambda}^{*},\bm{\nu}^{*})$ such that $g(\bm{\lambda}^{*},\bm{\nu}^{*})=d^{*}=p^{*}$, i.e., the optimal value of the primal problem can be obtained by solving the dual problem.

The Slater's condition holds also with a weaker constraint qualification, when some of the inequality constraint functions, $f_{1},\cdots,f_{k}$, are affine (instead of convex),
\begin{equation}
\begin{aligned}
& f_{i}(\mathbf{x})\leq 0, \; i = 1,\cdots,k, \\
& f_{i}(\mathbf{x})<0, \; i = k+1,\cdots,m, \\
& \mathbf{A}_{q\times N}\mathbf{x} = \mathbf{y}.
\end{aligned}
\label{eq:WeakSlatersCondition}
\end{equation}
The weaker constraint qualifications~\eqref{eq:WeakSlatersCondition} imply that strong duality reduces to feasibility when both the inequality and the equality constraints are linear.

\subsection{\label{sec:SchurComplement} Schur complement}

Let $\mathbf{X}$ be a square Hermitian matrix partitioned as,
\begin{equation}
\mathbf{X} = \begin{bmatrix} \mathbf{A} & \mathbf{B}\\
\mathbf{B}^{H} & \mathbf{C}\end{bmatrix},
\label{eq:HermitianPartitioned}
\end{equation}
\noindent where $\mathbf{A}$ is also square Hermitian. If $\det\mathbf{A}\neq 0$ then the matrix,
\begin{equation}
\mathbf{S} = \mathbf{C}-\mathbf{B}^{H}\mathbf{A}^{-1}\mathbf{B},
\label{eq:SchurComplement}
\end{equation}
\noindent is called the Schur complement of $\mathbf{A}$ in $\mathbf{X}$.

A useful property related to the Schur complement is that if $\mathbf{A}\succ 0$ then $\mathbf{X}\succeq 0$ if and only if $\mathbf{S}\succeq 0$.

\section{Bounded trigonometric polynomials}

This section presents useful results for bounded trigonometric polynomials and their roots as presented in Refs.~\onlinecite{DumitrescuBook}.

\subsection{Trigonometric polynomials}

Let $\mathbf{a}(\omega) =[1,e^{j\omega},\cdots,e^{j\omega (L-1)}]^{T}$ be a $L\times 1$ basis vector for trigonometric polynomials of degree $L-1$ with $\omega\in[-\pi,\pi]$. A (causal) trigonometric polynomial can be written in terms of the basis vector as,
\begin{equation}
H(\omega) = \sum\limits_{l=0}^{L-1} h_{l} e^{-j\omega l} = \mathbf{a}(\omega)^{H} \mathbf{h},
\label{eq:TrigonometricPolynomial}
\end{equation}
\noindent where $\mathbf{h} =\left[h_{0},\cdots,h_{L-1}\right]^{T} \in \mathbb{C}^{L}$ is the vector of the polynomial coefficients.

\subsection{Nonnegative trigonometric polynomials}

Let $R(\omega) = \lvert H(\omega)\rvert ^{2} = H(\omega) H(\omega)^{H}$. From~\eqref{eq:TrigonometricPolynomial}, the nonnegative trigonometric polynomial $R(\omega)$ has the form,
\begin{equation}
R(\omega) = \sum\limits_{k=-(L-1)}^{L-1} r_{k} e^{-j\omega k},
\label{eq:NNTP}
\end{equation}
\noindent where $r_{k} = \sum\limits_{l=0}^{L-1-k} h_{l}h_{l+k}^{*}$ for $k\geq 0$ and $r_{-k} = r_{k}^{*}$, i.e., the coefficients are conjugate symmetric thus $R(\omega)$ is Hermitian. Equivalently, the coefficients $r_{k}$ can be calculated as the sum of the $k^{th}$ diagonal elements of the autocorrelation matrix $\mathbf{Q}_{L\times L} = \mathbf{h}\mathbf{h}^{H}$ as,
\begin{equation}
r_{k}=\sum\limits_{i=1}^{L-k} \mathbf{Q}_{i,i+k}.
\label{eq:NNTPcoefficients}
\end{equation}

\subsection{Bounded trigonometric polynomials}

Let two polynomials $H(\omega)$ and $B(\omega)$ fulfill the inequality,
\begin{equation}
\lvert H(\omega)\rvert\leq \lvert B(\omega)\rvert, \; \forall \omega\in[-\pi, \pi], 
\label{eq:PolyBoundedPoly}
\end{equation}
\noindent which implies $\lvert H(\omega)\rvert^{2}\leq \lvert B(\omega)\rvert^{2}, \; \forall \omega\in[-\pi, \pi]$. Defining $R_{H}(\omega)=\lvert H(\omega)\rvert^{2}$ and $R_{B}(\omega)=\lvert B(\omega)\rvert^{2}$ as in~\eqref{eq:NNTP}, yields $R_{H}(\omega)\leq R_{B}(\omega)$. From Lemma 4.23 in \cite{DumitrescuBook}, $R_{H}(\omega)\leq R_{B}(\omega)$ implies $\mathbf{Q}_{H} \preceq \mathbf{Q}_{B}$, where $\mathbf{Q}_{H} = \mathbf{h}\mathbf{h}^{H}$ and $\mathbf{Q}_{B} = \mathbf{b}\mathbf{b}^{H}$ are the autocorrelation matrices of the coefficient  vectors $\mathbf{h} = [h_{0},\cdots, h_{L-1}]^{T}$ and $\mathbf{b} = [b_{0},\cdots, b_{L-1}]^{T}$ of the polynomials $H(\omega)$ and $B(\omega)$ respectively. Through a Schur complement (see Sec.~\ref{sec:SchurComplement}), $\mathbf{Q}_{B} - \mathbf{h}1^{-1}\mathbf{h}^{H}\succeq 0$ is equivalent to semidefinite matrix ,
\begin{equation}
\begin{bmatrix} \mathbf{Q}_{B} & \mathbf{h}_{L\times 1}\\
\mathbf{h}^{H}_{1\times L} & 1 \end{bmatrix} \succeq 0.
\label{eq:SchurConstraint}
\end{equation}

Let the polynomial $H(\omega)$ have amplitude uniformly bounded for all $\omega\in[-\pi, \pi]$ such that, $\lvert H(\omega)\rvert\leq \gamma$, where $\gamma \in \mathbb{R}_{+}$ is a given positive real number. As a special case of the results for bounded trigonometric polynomials in Eqs.~\eqref{eq:PolyBoundedPoly},~\eqref{eq:SchurConstraint}, with $\lvert B(\omega)\rvert=\gamma$, Theorem 4.24 and corollary 4.25 in \cite{DumitrescuBook} states that the inequality $\lvert H(\omega)\rvert\leq \gamma$ can be approximated by two linear matrix inequalities,
\begin{equation}
\begin{aligned}
& \begin{bmatrix} \mathbf{Q}_{L\times L} & \mathbf{h}_{L\times 1}\\
\mathbf{h}^{H}_{1\times L} & 1 \end{bmatrix} \succeq 0,\\
&\sum\limits_{i=1}^{L-j} \mathbf{Q}_{i,i+j} = \left\{ \begin{array}{rl}
 \gamma^{2}, &j=0 \\
 0, &j=1,\cdots,L-1.
       \end{array} \right.
\end{aligned}
\label{eq:BRL}
\end{equation}
\noindent The latter constraint follows from the autocorrelation matrix of the constant polynomial $R_{B}(\omega)=\gamma^{2}$.

The results for bounded trigonometric polynomials can be used in relation to the $\ell_{\infty}$-norm, since setting an upper bound for the maximum amplitude of a polynomial implies that the polynomial has amplitude uniformly bounded for all $\omega\in[-\pi, \pi]$, 
\begin{equation}
\begin{aligned}
&\lVert H \rVert_{\infty} = \underset{\omega\in[-\pi, \pi]}{\max} \lvert H(\omega) \rvert \leq \gamma,\\
& \lvert H(\omega) \rvert \leq \gamma, \; \forall \omega\in[-\pi, \pi].
\end{aligned}
\label{eq:BoundedLinfty}
\end{equation}

\subsection{\label{sec:Roots}Roots of real nonnegative trigonometric polynomials}

For a bounded trigonometric polynomial $\lvert H(\omega)\rvert\leq 1$, we can construct a polynomial,
\begin{equation}
P(\omega) = 1-\lvert H(\omega)\rvert^{2} = 1 - R(\omega),
\label{eq:RPTP}
\end{equation}
which is by definition real-valued and nonnegative, thus it cannot have single roots on the unit circle. The degree of the polynomial $P(\omega)$ is $2(L-1)$. Therefore, the polynomial $P(\omega)$ has at most $L-1$ distinct roots on the unit circle. At a root, $\omega_{0}$, we have $P(\omega_{0})=0$ and subsequently $\lvert H(\omega_{0})\rvert = 1$.

\section{\label{sec:MatlabCode} Implementation in Matlab}

%
%
%
%
%
%
%
%

The algorithm in Table~\ref{table:MatlabCode} for the implementation of the method described in Sec.~\ref{sec:NoiselessReconstruction} is an adaptation of the code by Fernandez-Granda in Ref. \onlinecite{GrandaSuperRes2014}.

\begin{table}[ht!]
\caption{Matlab code for Sec.~\ref{sec:NoiselessReconstruction}}
\centering
\begin{tabular}{l}
\hline\hline
\quad Given $\mathbf{y}\in \mathbb{C}^{M}$, $d$, $\lambda$\\
\quad Solve dual problem with CVX\cite{CVX}, Eq.~\eqref{eq:DualProblemSP}\\
\hline

1:  cvx\_solver sdpt3\\
2:  cvx\_begin sdp \\
3:\quad\;  variable $S(M+1,M+1)$ hermitian\\
4:\quad\;  $S >= 0$;\\
5:\quad\;  $S(M+1,M+1) == 1$;\\
6:\quad\;  trace($S$) $== 2$;\\
7:\quad\;  for $j = 1:M-1$\\
8:\quad\quad\;  sum(diag($S,j$)) $== S(M+1-j,M+1)$;\\
9:\quad\;   end\\
10:\quad  maximize(real($S(1:M,M+1)'*y$))\\
11:  cvx\_end\\
12: $c = S(1:M,M+1)$;\\

\hline
\quad Find the roots of $P_{+}$, Eq~\eqref{eq:RootPolyEquiv}\\
\hline

13: $r$ = conv($c$,flipud(conj($c$)));\\
14: $r(M) = 1 - r(M)$;\\
15: roots\_$P$ = roots($r$);\\

\hline
\quad Isolate roots on the unit circle, Eq.~\eqref{eq:SupportRecovery}\\
\hline

16: roots\_uc = roots\_$P$(abs(1-abs(roots\_$P$))$<1e-2$);\\
17: [aux,ind]=sort(real(roots\_uc));\\
18: roots\_uc = roots\_uc(ind);\\
19: $t$ = angle(roots\_uc($1$:$2$:end))/($2*pi*d/lambda$);\\

\hline
\quad Amplitude estimation, Eq.~\eqref{eq:AmplitudeEstimation}\\
\hline

20: A\_T = exp($1i*2*pi*d/lambda*$[$0$:$(M-1)$]$'*t'$);\\
21: x\_CS\_dual = A\_T$\backslash y;$ \\

\hline\hline
\end{tabular}
\label{table:MatlabCode}
\end{table}

\section{\label{sec:AppDualNoise} Dual problem with noise}

In the case that the measurements~\eqref{eq:NoisyDataContinuous} are contaminated with additive noise $\mathbf{n}\in \mathbb{C}^{M}$ such that $\lVert\mathbf{n}\rVert_{2}\leq \epsilon$, the primal problem of atomic norm minimization~\eqref{eq:PrimalProblem} is reformulated to the problem~\eqref{eq:PrimalProblemNoise} or equivalently,
\begin{equation}
\underset{x}{\min}\lVert x\rVert_{\mathcal{A}} \; \text{subject to} \; 
\left\{\begin{array}{lr}
 \mathbf{y} =  \mathcal{F}_{M}x +\mathbf{n},\\
 \lVert \mathbf{n} \rVert_{2}\leq \epsilon.
\end{array}\right.
\label{eq:PrimalProblemNoise2Constr}
\end{equation}

The Lagrangian for~\eqref{eq:PrimalProblemNoise2Constr} is formulated by augmenting the objective function with a weighted sum of the constraints,
\begin{equation}
\begin{aligned}
&L(x,\mathbf{c},\xi) =\\
& \lVert x\rVert_{\mathcal{A}} + \mathrm{Re}\left[ \mathbf{c}^{H}\left( \mathbf{y}-\mathcal{F}_{M}x-\mathbf{n}\right)\right] + \xi\left(\mathbf{n}^{H}\mathbf{n}-\epsilon^{2}\right),
\end{aligned}
\label{eq:LagrangianNoise}
\end{equation}
\noindent where $\mathbf{c}\in\mathbb{C}^{M}$ are the dual variables related to the equality constraints, $\mathbf{y} - \mathcal{F}_{M}x -\mathbf{n} =0$, and $\xi\in\mathbb{R}^{+}$ is a Lagrange multiplier related to the inequality constraint, $\lVert \mathbf{n} \rVert_{2}-\epsilon\leq 0$.

The dual function $g(\mathbf{c},\xi)$ is the  infimum of the Lagrangian, $L(x,\mathbf{c},\xi)$, over the optimization variable x, 
\begin{equation}
\begin{aligned}
g(\mathbf{c},\xi) &= \underset{x}{\inf} \; L(x,\mathbf{c},\xi)\\ 
&= \mathrm{Re}\left[ \mathbf{c}^{H}\mathbf{y}-\mathbf{c}^{H}\mathbf{n}\right] + \xi\left(\mathbf{n}^{H}\mathbf{n}-\epsilon^{2}\right)+\\
&+\underset{x}{\inf} \; \left( \lVert x\rVert_{\mathcal{A}} - \mathrm{Re}\left[\mathbf{c}^{H}\mathcal{F}_{M}x\right] \right).
\end{aligned}
\label{eq:DualFunctionNoise}
\end{equation}
\noindent Minimizing over the unknown noise $\mathbf{n}\in\mathbb{C}^{M}$,
\begin{equation}
\frac{\partial g(\mathbf{c},\xi)}{\partial{\mathbf{n}}} = -\mathbf{c} +2\xi\mathbf{n} = 0,
\label{eq:DualFunctionDerivN}
\end{equation}
\noindent yields the optimal noise vector, $\mathbf{n}_{\mathrm{o}} = \mathbf{c}/\left( 2\xi\right)$. The dual function evaluated at $\mathbf{n}_{\mathrm{o}}$ is,
\begin{equation}
\begin{aligned}
g(\mathbf{c},\xi)|_{\mathbf{n}_{\mathrm{o}}} &= \mathrm{Re}\left[ \mathbf{c}^{H}\mathbf{y}\right]-\frac{\mathbf{c}^{H}\mathbf{c}}{2\xi} + \xi\left(\frac{\mathbf{c}^{H}\mathbf{c}}{4\xi^{2}}-\epsilon^{2}\right)+\\
&+\underset{x}{\inf} \; \left( \lVert x\rVert_{\mathcal{A}} - \mathrm{Re}\left[\mathbf{c}^{H}\mathcal{F}_{M}x\right] \right).
\end{aligned}
\label{eq:DualFunctionNoiseNo}
\end{equation}
\noindent Further, maximizing over the dual variable $\xi$,
\begin{equation}
\frac{\partial g(\mathbf{c},\xi)|_{\mathbf{n}_{\mathrm{o}}}}{\partial{\xi}} = \frac{\mathbf{c}^{H}\mathbf{c}}{4\xi^{2}}-\epsilon^{2} = 0,
\label{eq:DualFunctionDerivXi}
\end{equation}
\noindent we obtain the optimal value for the dual variable $\xi_{\mathrm{o}} = \lVert \mathbf{c}\rVert_{2}/\left( 2\epsilon\right)$.

Finally, the dual function evaluated at the optimal values $\mathbf{n}_{\mathrm{o}}$ and $\xi_{\mathrm{o}}$ becomes,
\begin{equation}
\begin{aligned}
g(\mathbf{c})|_{\mathbf{n}_{\mathrm{o}},\xi_{\mathrm{o}}} &= \mathrm{Re}\left[ \mathbf{c}^{H}\mathbf{y}\right]-\epsilon\lVert\mathbf{c}\rVert_{2}+\\
&+\underset{x}{\inf} \; \left( \lVert x\rVert_{\mathcal{A}} - \mathrm{Re}\left[\mathbf{c}^{H}\mathcal{F}_{M}x\right] \right),
\end{aligned}
\label{eq:DualFunctionNoiseNoXo}
\end{equation}
\noindent and the dual problem is formulated by maximizing the dual function, $g(\mathbf{c})|_{\mathbf{n}_{\mathrm{o}},\xi_{\mathrm{o}}}$, over the dual variables $\mathbf{c}\in\mathbb{C}^{M}$ similarly to the process detailed in Sec.~\ref{sec:DualProblem},
\begin{equation}
\begin{aligned}
&\underset{\mathbf{c}}{\max} \; g(\mathbf{c})|_{\mathbf{n}_{\mathrm{o}},\xi_{\mathrm{o}}} \equiv\\
&\underset{\mathbf{c}}{\max}\;\mathrm{Re}\left[\mathbf{c}^{H}\mathbf{y}\right] -\epsilon\lVert\mathbf{c}\rVert_{2}  \; \text{subject to} \;  \lVert\mathcal{F}^{H}_{M}\mathbf{c}\rVert_{\infty}\leq 1.
\end{aligned}
\label{eq:DualProblemNoiseApp}
\end{equation}


\begin{thebibliography}{}
\newcommand{\enquote}[1]{``#1''}
\expandafter\ifx\csname url\endcsname\relax
  \def\url#1{\texttt{#1}}\fi
\expandafter\ifx\csname urlprefix\endcsname\relax\def\urlprefix{URL }\fi
\providecommand{\bibinfo}[2]{#2}
\providecommand{\noopsort}[1]{}
\providecommand{\switchargs}[2]{#2#1}

\end{thebibliography}


\begin{thebibliography}{10}

\bibitem{EladBook:2010}
M.~Elad.
\newblock {\em Sparse and redundant representations: from theory to
  applications in signal and image processing}, pages 1--359.
\newblock Springer, New York, 2010.

\bibitem{FoucartBook:2013}
S.~Foucart and H.~Rauhut.
\newblock {\em A mathematical introduction to compressive sensing}, pages
  1--589.
\newblock Springer, New York, 2013.

\bibitem{MalioutovDOA:2005}
D.~Malioutov, M.~{\c{C}}etin, and A.~S. Willsky.
\newblock A sparse signal reconstruction perspective for source localization
  with sensor arrays.
\newblock {\em IEEE Trans. Signal Process.}, \textbf{53}(8):3010--3022, 2005.

\bibitem{EdelmannCSDOA:2011}
G.~F. Edelmann and C.~F. Gaumond.
\newblock Beamforming using compressive sensing.
\newblock {\em J. Acoust. Soc. Am.}, \textbf{130}(4):232--237, 2011.

\bibitem{XenakiCS:2014}
A.~Xenaki, P.~Gerstoft, and K.~Mosegaard.
\newblock Compressive beamforming.
\newblock {\em J. Acoust. Soc. Am.}, \textbf{136}(1):260--271, 2014.

\bibitem{Mecklenbrauker:2013}
C.~F. Mecklenbr\"auker, P.~Gerstoft, A.~Panahi, and M.~Viberg.
\newblock Sequential {B}ayesian sparse signal reconstruction using array data.
\newblock {\em IEEE Trans. Signal Process.}, \textbf{61}(24):6344--6354, 2013.

\bibitem{KrimDOA:1996}
H.~Krim and M.~Viberg.
\newblock Two decades of array signal processing research: the parametric
  approach.
\newblock {\em IEEE Signal Proc. Mag.}, \textbf{13}(4):67--94, 1996.

\bibitem{MantzelCMFP:2012}
W.~Mantzel, J.~Romberg, and K.~Sabra.
\newblock Compressive matched-field processing.
\newblock {\em J. Acoust. Soc. Am.}, \textbf{132}(1):90--102, 2012.

\bibitem{Forero:2014}
P.~A. Forero and P.~A. Baxley.
\newblock Shallow-water sparsity-cognizant source-location mapping.
\newblock {\em J. Acoust. Soc. Am.}, \textbf{135}(6):3483--3501, 2014.

\bibitem{YardimCSfathometer:2014}
C.~Yardim, P.~Gerstoft, W.~S. Hodgkiss, and Traer J.
\newblock Compressive geoacoustic inversion using ambient noise.
\newblock {\em J. Acoust. Soc. Am.}, \textbf{135}(3):1245--1255, 2014.

\bibitem{ChiBasisMismatch:2011}
Y.~Chi, L.~L. Scharf, A.~Pezeshki, and A.~R. Calderbank.
\newblock Sensitivity to basis mismatch in compressed sensing.
\newblock {\em IEEE Trans. Signal Process.}, \textbf{59}(5):2182--2195, 2011.

\bibitem{DuarteBasisMismatch:2013}
M.~F. Duarte and R.~G. Baraniuk.
\newblock Spectral compressive sensing.
\newblock {\em Appl. Comput. Harmon. Anal.}, \textbf{35}(1):111--129, 2013.

\bibitem{YaoCSseismic:2011}
H.~Yao, P.~Gerstoft, P.~M. Shearer, and C.~Mecklenbr{\"a}uker.
\newblock Compressive sensing of the {T}ohoku-{O}ki {M}w 9.0 earthquake:
  {F}requency-dependent rupture modes.
\newblock {\em Geophys. Res. Lett.}, \textbf{38}(20):1--5, 2011.

\bibitem{YaoCSseismic:2013}
H.~Yao, P.~M. Shearer, and P.~Gerstoft.
\newblock Compressive sensing of frequency-dependent seismic radiation from
  subduction zone megathrust ruptures.
\newblock {\em Proc. Natl. Acad. Sci. U.S.A.}, \textbf{110}(12):4512--4517,
  2013.

\bibitem{Fan:2014}
W.~Fan, P.~M. Shearer, and P.~Gerstoft.
\newblock Kinematic earthquake rupture inversion in the frequency domain.
\newblock {\em Geophys. J. Int.}, \textbf{199}(2):1138--1160, 2014.

\bibitem{Chandrasekaran2012}
V.~Chandrasekaran, B.~Recht, P.~A. Parrilo, and A.~S. Willsky.
\newblock The convex geometry of linear inverse problems.
\newblock {\em Found. Comput. Math.}, \textbf{12}(6):805--849, 2012.

\bibitem{GrandaSuperRes2014}
E.~J. Cand{\`e}s and C.~Fernandez-Granda.
\newblock Towards a mathematical theory of super-resolution.
\newblock {\em Comm. Pure Appl. Math.}, \textbf{67}(6):906--956, 2014.

\bibitem{VanTreesBook}
H.L. Van~Trees.
\newblock {\em Optimum Array Processing (Detection, Estimation, and Modulation
  Theory, Part IV)}, chapter 1--10.
\newblock Wiley-Interscience, New York, 2002.

\bibitem{FORAdata:2011}
H.~C. Song, S.~Cho, T.~Kang, W.~S. Hodgkiss, and J.~R. Preston.
\newblock Long-range acoustic communication in deep water using a towed array.
\newblock {\em J. Acoust. Soc. Am.}, \textbf{129}(3):71--75, 2011.

\bibitem{JohnsonBook1993}
D.~H. Johnson and D.~E. Dudgeon.
\newblock {\em Array signal processing: concepts and techniques}, pages 1--512.
\newblock PRT Prentice Hall, Englewood Cliffs, NJ, 1993.

\bibitem{BaraniukCSNotes:2007}
R.~G. Baraniuk.
\newblock Compressive sensing.
\newblock {\em IEEE Signal Proc. Mag.}, \textbf{24}(4):118--121, 2007.

\bibitem{TroppCSNoise:2006}
J.~A. Tropp.
\newblock Just relax: Convex programming methods for identifying sparse signals
  in noise.
\newblock {\em IEEE Trans. Inf. Theory}, \textbf{52}(3):1030--1051, 2006.

\bibitem{CVX}
M.~Grant and S.~Boyd.
\newblock {CVX}: {M}atlab software for disciplined convex programming, version
  2.0 beta.
\newblock \url{http://cvxr.com/cvx}, September 2013.

\bibitem{BoydBook}
S.~Boyd and L.~Vandenberghe.
\newblock {\em Convex optimization}, pages 1--684.
\newblock Cambridge university press, New York, 2004.

\bibitem{CandesRIP:2008}
E.~J. Cand{\`e}s.
\newblock The restricted isometry property and its implications for compressed
  sensing.
\newblock {\em C. R. Math. Acad. Sci.}, \textbf{346}(9):589--592, 2008.

\bibitem{Fuchs:2005}
J.~J. Fuchs.
\newblock Sparsity and uniqueness for some specific under-determined linear
  systems.
\newblock In {\em IEEE International Conference on Acoustics, Speech, and
  Signal Processing, ICASSP'05}, volume \textbf{5}, pages 729--732. IEEE, 2005.

\bibitem{TangCSoffGrid013}
G.~Tang, B.~N. Bhaskar, P.~Shah, and B.~Recht.
\newblock Compressed sensing off the grid.
\newblock {\em IEEE Trans. Inf. Theory}, \textbf{59}(11):7465--7490, 2013.

\bibitem{GrandaNoisyData2013}
E.~J. Cand{\`e}s and C.~Fernandez-Granda.
\newblock Super-resolution from noisy data.
\newblock {\em J. Fourier Anal. Appl.}, \textbf{19}(6):1229--1254, 2013.

\bibitem{Capon:1969}
J.~Capon.
\newblock High-resolution frequency-wavenumber spectrum analysis.
\newblock {\em Proc. IEEE}, \textbf{57}(8):1408--1418, 1969.

\bibitem{SchmidtMUSIC:1986}
R.~Schmidt.
\newblock Multiple emitter location and signal parameter estimation.
\newblock {\em IEEE Trans. Antennas Propag.}, \textbf{34}(3):276--280, 1986.

\bibitem{KumaresanMinNorm:1983}
R.~Kumaresan and D.~W. Tufts.
\newblock Estimating the angles of arrival of multiple plane waves.
\newblock {\em IEEE Trans. Aerosp. Electron. Syst.}, \textbf{19}(1):134--139,
  1983.

\bibitem{KumaresanZeros:1983}
R.~Kumaresan.
\newblock On the zeros of the linear prediction-error filter for deterministic
  signals.
\newblock {\em IEEE Trans. Acoust., Speech, Signal Process.},
  \textbf{31}(1):217--220, 1983.

\bibitem{Barabell:1983}
A.~Barabell.
\newblock Improving the resolution performance of eigenstructure-based
  direction-finding algorithms.
\newblock In {\em IEEE International Conference on Acoustics, Speech, and
  Signal Processing, ICASSP'83}, volume \textbf{8}, pages 336--339. IEEE, 1983.

\bibitem{Rao:1989}
B.~D. Rao and K.~V.~S. Hari.
\newblock Performance analysis of root-{MUSIC}.
\newblock {\em IEEE Trans. Acoust., Speech, Signal Process.},
  \textbf{37}(12):1939--1949, 1989.

\bibitem{Pillai:1989}
S.~U. Pillai and B.~H. Kwon.
\newblock Forward/backward spatial smoothing techniques for coherent signal
  identification.
\newblock {\em IEEE Trans. Acoust., Speech, Signal Process.},
  \textbf{37}(1):8--15, 1989.

\bibitem{RaoSpatialSmoothing:1990}
B.~D. Rao and K.~V.~S. Hari.
\newblock Effect of spatial smoothing on the performance of {MUSIC} and the
  minimum-norm method.
\newblock {\em IEE Proc. Radar and Signal Proces.}, \textbf{137}(6):449--458,
  1990.

\bibitem{DumitrescuBook}
B.~Dumitrescu.
\newblock {\em Positive trigonometric polynomials and signal processing
  applications}, chapter 4.3.
\newblock Springer, Dordrecht, Netherlands, 2007.

\end{thebibliography}
\end{document}